\documentclass[a4paper,12pt,final,reqno]{amsart}

\usepackage{amsmath,amssymb,amsthm}

\numberwithin{equation}{section}
\allowdisplaybreaks

\theoremstyle{definition}
 \newtheorem{thm}{Theorem}[section]
 \newtheorem{prp}[thm]{Proposition}

 \newtheorem{con}[thm]{Conjecture}
 \newtheorem{fct}[thm]{Fact}
 \newtheorem{dfn}[thm]{Definition}
 \newtheorem{eg}[thm]{Example}

 \newtheorem{rmk}[thm]{Remark}

\newcommand{\bra}[1]{\left< #1 \right|}
\newcommand{\ket}[1]{\left|#1 \right>}
\newcommand{\seteq}{\mathbin{:=}}

\newcommand{\id}{\operatorname{id}}

\newcommand{\Hilb}{\operatorname{Hilb}\nolimits}

\newcommand{\SU}{\operatorname{SU}\nolimits}
\newcommand{\U}{\operatorname{U}\nolimits}

\newcommand{\Vir}{\operatorname{Vir}\nolimits}
\newcommand{\simto}{\xrightarrow{\,\sim\,}}
\renewcommand{\Vec}[1]{{\boldsymbol #1}}

\newcommand{\ve}{\varepsilon}
\newcommand{\bbC}{\mathbb{C}}
\newcommand{\bbF}{\mathbb{F}}

\newcommand{\bbP}{\mathbb{P}}
\newcommand{\bbQ}{\mathbb{Q}}
\newcommand{\bbT}{\mathbb{T}}
\newcommand{\bbZ}{\mathbb{Z}}
\newcommand{\bfal}{\boldsymbol{\alpha}}
\newcommand{\bfbe}{\boldsymbol{\beta}}
\newcommand{\bfu}{\mathbf{u}}
\newcommand{\bfv}{\mathbf{v}}
\newcommand{\bfw}{\mathbf{w}}
\newcommand{\calA}{\mathcal{A}}

\newcommand{\calF}{\mathcal{F}}
\newcommand{\calH}{\mathcal{H}}

\newcommand{\calP}{\mathcal{P}}
\newcommand{\calU}{\mathcal{U}}
\newcommand{\calW}{\mathcal{W}}
\newcommand{\frkg}{\mathfrak{g}}
\newcommand{\frkn}{\mathfrak{n}}

\newcommand{\FF}{\widetilde{\bbF}}
\newcommand{\va}{\overrightarrow{a}}
\newcommand{\vlambda}{\Vec{\lambda}}
\newcommand{\vmu}{\Vec{\mu}}

\title{Notes on Ding-Iohara algebra and AGT conjecture }
\author{H.~Awata, B.~Feigin, A.~Hoshino, M.~Kanai, J.~Shiraishi and S.~Yanagida}
\address{HA: Graduate School of Mathematics, Nagoya University, Nagoya, 464-8602, Japan}
\email{awata@math.nagoya-u.ac.jp}
\address{BF: Landau Institute for Theoretical Physics,
Russia, Chernogolovka, 142432, prosp. Akademika Semenova, 1a,   
\\
Higher School of Economics, Russia, Moscow, 101000,  Myasnitskaya ul., 20, and
\\
Independent University of Moscow, Russia, Moscow, 119002,
Bol'shoi Vlas'evski per., 11}
\email{bfeigin@gmail.com}
\address{AH. Department of Mathematics, Sophia University, Kioicyo, Tokyo, 102-8554, Japan}
\email{ayumu-h@sophia.ac.jp}
\address{MK,JS: Graduate School of Mathematical Sciences, University of Tokyo, Komaba, Tokyo
153-8914, Japan}
\email{kanai@ms.u-tokyo.ac.jp}
\email{shiraish@ms.u-tokyo.ac.jp}
\address{SY: Kobe University, Department of Mathematics, Rokko, Kobe 657-8501, Japan}
\email{yanagida@math.kobe-u.ac.jp}

\begin{document}

\begin{abstract}
We study the representation theory of the 
Ding-Iohara algebra $\calU$ to find $q$-analogues of 
the Alday-Gaiotto-Tachikawa (AGT) relations.
We introduce
the endomorphism $T(u,v)$  of the Ding-Iohara algebra, 
having two parameters $u$ and $v$. 
We define the vertex operator $\Phi(w)$ 
by  specifying the permutation relations with the Ding-Iohara generators 
$x^\pm(z)$ and $\psi^\pm(z)$ in terms of  $T(u,v)$.
For the level one representation, 
all the matrix elements of the vertex operators
with respect to the Macdonald polynomials are factorized and written 
in terms of  the Nekrasov factors 
for the $K$-theoretic partition functions as in the AGT relations.
For higher levels $m=2,3,\ldots$, 
we present some conjectures, which imply the existence of the $q$-analogues of the 
AGT relations.

\end{abstract}

\maketitle

%%%%%%%%%%%%%%%%%%%%%%%%%%%%
\section{Introduction}
\label{Intro}

%%%%%%%%%%%%%%%%%%%%%%%%%%%%%%
The aim of this note is to continue our study on the 
representation theory of the Ding-Iohara algebra $\calU$ \cite{DI:1997}
on positive integer levels, and to search a connection with the 
findings of Alday, Gaiotto and Tachikawa (AGT) \cite{AGT:2010}.
Authors' previous discussions on $\calU$ 
are found in  \cite{FHHSY:2009} and  \cite{FHSSY:2010}.
As for the related works, see \cite{FT:2009}, \cite{SV:2009}, 
\cite{FFJMM:2010-1}, \cite{FFJMM:2010-2} and \cite{S:2010}.
\\

In  \cite{FHHSY:2009},
we studied the level one action of the Ding-Iohara algebra 
$\calU$ on the space of Macdonald symmetric 
functions $P_\lambda(x;q,t)$, namely on the Fock space $\calF_u$
(see \S \ref{DI}, \S \ref{Mac} and \S \ref{level_one_Fock}).
In  \cite{FHSSY:2010}, we showed that for positive
integer levels $m=2,3,\ldots$, the  Ding-Iohara
algebra is realized on the $m$-fold tensor space 
$\calF_{u_1}\otimes \calF_{u_2}\otimes \cdots \otimes \calF_{u_m}$
by  the deformed $\calW_m$ algebra together with an extra 
Heisenberg algebra. 
In this note, we introduce several bases on the 
$m$-fold tensor representation space. 
The first is the Macdonald-type basis $(\ket{P_\vlambda})$ (see \S \ref{Mac_level_m}). 
Here $\vlambda=(\lambda^{(1)},\lambda^{(2)},\ldots,\lambda^{(m)})$,
and each component $\lambda^{(i)}$ is a partition.
Next we introduce the `Poincar\'e-Birkoff-Witt-type basis' $(\ket{X_\vlambda})$,
and the `integral basis' 
$(\ket{K_\vlambda})$ (see \S \ref{K_lambda:level_one} and \S \ref{K_lambda:level_m}).
%Introducing a Poincar\'e-Birkoff-Witt-type basis.
In the level one case, we  can show that  $(\ket{K_\lambda})$ 
essentially gives the integral form  $J_\lambda(x;q,t)$ (see Proposition \ref{prop:level1:K}).
Unfortunately, at this moment, we do not have proofs that 
$(\ket{X_\vlambda})$ and $(\ket{K_\lambda})$ are bases for higher level cases $m=2,3,\ldots$.
\medskip

We introduce an
endomorphism 
$T(u,v)$ acting on the Ding-Iohara algebra having two parameters $u$ and $v$ 
(see Definition \ref{Tuv}).
In the level one case, we define the vertex operator $\Phi(w):\calF_u\rightarrow \calF_v$ 
by the normalization $\Phi(w)\ket{0}=\ket{0}+\cdots $, 
and  the permutation relations
$T(vw,q^{-1}t uw)(a)\Phi(w)
=\Phi(w)T(q^{-1}t vw,uw)(a)$ for all $ a\in \calU$ (see Definition \ref{defVOlevelone}).
Then we claim that 
\begin{enumerate}
\item (Proposition \ref{Phi=}) the $\Phi(w)$ exists uniquely as
\begin{align*}
&\Phi(w) =
 \exp \Bigl(
 -\sum_{n=1}^{\infty} \dfrac{1}{n}\dfrac{v^n-(t/q)^{n}u^n}{1-q^n} a_{-n}w^n 
      \Bigr)
 \exp\Bigl(
  \sum_{n=1}^{\infty} \dfrac{1}{n}\dfrac{v^{-n}- u^{-n}}{1-q^{-n}} a_{n}w^{-n}
     \Bigr),
\end{align*}
\item  (Proposition \ref{conj:level1}) 
all the matrix elements $\bra{K_\lambda}\Phi(w)\ket{K_\mu}$
are factorized as
\begin{align*}
\bra{  K_\lambda} \Phi(w) \ket{K_\mu} =
N_{\lambda,\mu}(q v/t u)
(-t u v w/q)^{|\lambda|}(t v w/q)^{-|\mu|} 
u^{|\mu|} t^{-n(\mu)}q^{n(\mu')}.
\end{align*}

 \end{enumerate}
Here $a_n$'s denote the Heisenberg generators satisfying 
$[a_m,a_n]=\delta_{m+n,0} m (1-q^{|m|})/(1-t^{|m|})$,
and we have used the notation for 
the `$K$-theoretic Nekrasov factor' (see Definition \ref{dfn:N})
\begin{align*}
N_{\lambda,\mu}(u)
&\seteq
\prod_{(i,j)\in \lambda}(1-u q^{-\mu_i+j-1}t^{-\lambda'_j+i}) \cdot 
\prod_{(k,l)\in \mu}(1-u q^{\lambda_k-l}t^{\mu'_l-k+1}) \\
&=
\prod_{\square\in \lambda}
(1-u q^{-a_\mu(\square)-1}t^{-\ell_\lambda(\square)}) \cdot 
\prod_{\blacksquare \in \mu}
(1-u q^{a_\lambda(\blacksquare)}t^{\ell_\mu(\blacksquare)+1}).
\end{align*}
See \S \ref{Mac} for the combinatorial symbols used here.
Hence we found a $q$-analogue of the 
AGT relation  \cite{AGT:2010} for the case the gauge group is $\U(1)$.
\medskip

For higher level cases, 
we define the vertex operator $\Phi(w)$ in a similar manner 
(see  Definition \ref{defVOlevelm}). 
Then we present our main conjecture about the properties of $ \Phi(w)$ 
(see Conjecture \ref{maincon}).
Our conjecture implies that we have $q$-deformed AGT relation
for the case the gauge group is $\U(m)$.
\medskip

This note is organised as follows.
In Section 2, we recall the definition of the Ding-Iohara algebra $\calU$, 
the Macdonald polynomials, and the level one representation of $\calU$ on the
Fock space $\calF_u$. We give the definitions of the integral basis $\ket{K_\lambda}$ 
and the vertex operator $\Phi(w)$. Then we state the properties of $\Phi(w)$
in Proposition \ref{conj:level1}. In Section 3, 
we study the level $m$ representation given on the 
$m$-fold tensor space 
$\calF_{u_1}\otimes\calF_{u_2}\otimes\cdots \otimes \calF_{u_m}$. 
In Conjecture \ref{maincon}, we summarize our observation about 
the vertex operator $\Phi(w)$.
Section 4 is devoted to a brief review of the AGT conjecture, 
Whittaker or Gaiotto state, and their five dimensional version.
In Section 5, we study the Whittaker vectors for the Ding-Iohara algebra.
In Section 6, we give some examples of 
calculating the matrix elements of $\Phi(w)$ for the level one case.

%%%%%%%%%%%%%%%%%%%%%%%%%
\section{Level One Representation}\label{Section_one}
\subsection{Ding-Iohara algebra}
\label{DI}
Recall the Ding-Iohara algebra \cite{DI:1997}.
Let $q,t$ be independent indeterminates and $\bbF\seteq\bbQ(q,t)$. 
Let $g(z)$ be the formal series 
\begin{align*}
g(z)\seteq\dfrac{G^+(z)}{G^-(z)}\in\bbF[[z]],\qquad
G^\pm(z)\seteq(1-q^{\pm1}z)(1-t^{\mp 1}z)(1-q^{\mp1}t^{\pm 1}z).
\end{align*}
We have $g(z)=g(z^{-1})^{-1}$
as is required.
\begin{dfn}
Let $\calU$ be the unital associative algebra over $\bbF$ 
generated by the Drinfeld currents 
$x^\pm(z)=\sum_{n\in \bbZ}x^\pm_n z^{-n}$,
$\psi^\pm(z)=\sum_{\pm n\in \bbZ_{\ge0}}\psi^\pm_n z^{-n}$
and the central element $\gamma^{\pm 1/2}$, satisfying the defining relations
\begin{align*}
&\psi^\pm(z) \psi^\pm(w) = \psi^\pm(w) \psi^\pm(z),
 \qquad\qquad\qquad\qquad
 \psi^+(z)\psi^-(w) =
 \dfrac{g(\gamma^{+1} w/z)}{g(\gamma^{-1}w/z)}\psi^-(w)\psi^+(z),
\\
&\psi^+(z)x^\pm(w) = g(\gamma^{\mp 1/2}w/z)^{\mp1} x^\pm(w)\psi^+(z),
 \qquad
 \psi^-(z)x^\pm(w) = g(\gamma^{\mp 1/2}z/w)^{\pm1} x^\pm(w)\psi^-(z),
\\
&[x^+(z),x^-(w)]
 =\dfrac{(1-q)(1-1/t)}{1-q/t}
 \big( \delta(\gamma^{-1}z/w)\psi^+(\gamma^{1/2}w)-
 \delta(\gamma z/w)\psi^-(\gamma^{-1/2}w) \big),
\\
&G^{\mp}(z/w)x^\pm(z)x^\pm(w)=G^{\pm}(z/w)x^\pm(w)x^\pm(z).
\end{align*}
\end{dfn}
\begin{fct}
The algebra $\calU$ has a formal Hopf algebra structure.
The formulas for the coproduct read
$\Delta(\gamma^{\pm 1/2})=\gamma^{\pm 1/2} \otimes \gamma^{\pm 1/2}$ and
\begin{align*}
&\Delta (\psi^\pm(z))=
 \psi^\pm (\gamma_{(2)}^{\pm 1/2}z)\otimes \psi^\pm (\gamma_{(1)}^{\mp 1/2}z),
\\
&\Delta (x^+(z))=
  x^+(z)\otimes 1+
  \psi^-(\gamma_{(1)}^{1/2}z)\otimes x^+(\gamma_{(1)}z),\\
&\Delta (x^-(z))=
  x^-(\gamma_{(2)}z)\otimes \psi^+(\gamma_{(2)}^{1/2}z)+1 \otimes x^-(z),
\end{align*}
where $\gamma_{(1)}^{\pm 1/2} \seteq \gamma^{\pm 1/2}\otimes 1$
and   $\gamma_{(2)}^{\pm 1/2} \seteq 1\otimes \gamma^{\pm 1/2}$.
Since we do not use the antipode $a$ and the counit $\varepsilon$ in this paper, we omit them.
\end{fct}

When the central element takes the value $\gamma^{\pm 1/2}=(t/q)^{\pm m/4}$
on a representation space with some $m\in \bbQ$,
we call it of {\it level $m$}.
\medskip

Now we introduce our main tool in the present paper.
\begin{dfn}\label{Tuv}
For generic parameters $u$ and $v$,
define the endomorphism $T(u,v)$ of $\calU$  by
\begin{align*}
&T(u,v) (x^+(z))=(1-u/z)x^+(z), \\
& T(u,v)  (x^-(z))=(1-\gamma v/z)x^-(z), \\
&T(u,v) (\psi^\pm(z))=
(1-\gamma^{\mp 1/2} u/z)(1-\gamma^{1\pm 1/2} v/z)\psi^\pm(z) ,
\end{align*}
where $\gamma$ is the central element. 
In Fourier modes, we have
\begin{align*}
&T(u,v) (x^+_n)=x^+_n-u x^+_{n-1},\\
&T(u,v) (x^-_n)=x^+_n-\gamma v x^-_{n-1},\\
&T(u,v) (\psi^\pm_n)=\psi^\pm_n-
(\gamma^{\mp 1/2} u+\gamma^{1\pm 1/2} v)\psi^\pm_{n-1}+
\gamma u v \psi^\pm_{n-2}.
\end{align*}
\end{dfn}

The endomorphism $T(u,v)$ will be used 
for giving the defining relations for our vertex operator  $\Phi(w)$.
See  Definition \ref{defVOlevelone} and Definition \ref{defVOlevelm} below.

\begin{rmk}
The image $T(u,v) (\calU)$ is strictly smaller than $\calU$.
Formally we can write $T(u,v)^{-1} (x^+_n)=
x^+_n+u x^+_{n-1}+u^{2}x^+_{n-2}+\cdots$ but 
this does not belong to $\calU$ because of the infinite sum. 
It might be an interesting problem to 
find some meaning to the formal inverse, however, we will not 
consider it in this paper.
\end{rmk}

%%%%%%%%%%%%%%%%%%%%%%%
\subsection{Macdonald polynomials}
\label{Mac}
We basically follow \cite{M:1995} for the notations. 
A partition $\lambda$ is a series of 
nonnegative integers $\lambda=(\lambda_1,\lambda_2,\ldots)$ 
such that $\lambda_1\ge\lambda_2\ge\cdots$ with finitely many nonzero entries.
We use the following symbols:
$|\lambda|  \seteq \sum_{i\geq 1} \lambda_i$, 
$n(\lambda) \seteq \sum_{i\geq 1}(i-1)\lambda_i$. If $\lambda_l>0$ and $\lambda_{l+1}=0$,
we write $\ell(\lambda) \seteq l$ and call it  the length of $\lambda$.
The conjugate partition of $\lambda$ is denoted by $\lambda'$ which corresponds to 
the transpose of the diagram $\lambda$.
The empty sequence is denoted by $\emptyset$.
The dominance ordering is defined by $\lambda\ge\mu$ $\Leftrightarrow$
$|\lambda|=|\mu|$ and 
$\sum_{k=1}^i \lambda_k \ge \sum_{k=1}^i \mu_k$ for all $i=1,2,\ldots$.

We also follow \cite{M:1995} for the convention of the Young diagram.
Namely, the first coordinate $i$ (the row index) increases as one goes downwards,
and the second coordinate $j$ (the column index) increases 
as one goes rightwards. 
We denote by $\square=(i,j)$ the box located at the coordinate $(i,j)$.
For a box $\square=(i,j)$ and a partition $\lambda$, we use the following notations:
\begin{align*}
i(\square)\seteq i,\quad 
j(\square)\seteq j,\quad
a_{\lambda}(\square)\seteq \lambda_i-j,\quad
\ell_{\lambda}(\square)\seteq \lambda'_j-i. %\quad
%a'_{\lambda}(\square)\seteq j-1,\quad
%\ell'_{\lambda}(\square)\seteq i-1.
\end{align*}

Let $\Lambda$ be the ring of symmetric functions in $x=(x_1,x_2,\ldots)$ over $\bbZ$,
and let $\Lambda_{\bbF}\seteq\Lambda\otimes_{\bbZ}\bbF$. 
Let $m_\lambda$ be the monomial symmetric functions.
Denote the power sum function by $p_n=\sum_{i\geq 1}x_i^n$. For a partition $\lambda$, 
we write
$p_\lambda=\prod_i p_{\lambda_i}$.
Macdonald's scalar product on $\Lambda_{\bbF}$ is
\begin{align}
\langle p_\lambda,p_\mu \rangle_{q,t}=\delta_{\lambda,\mu}
z_\lambda \prod_{i=1}^{\ell(\lambda)} {1-q^{\lambda_i}\over 1-t^{\lambda_i}},\qquad 
z_\lambda=\prod_{i\geq 1} i^{m_i} \cdot m_i!, \label{eq:Macdpair}
\end{align}
Here we denote by $m_i$ 
the number of entries  in $\lambda$ equal to $i$.
\begin{fct}
The Macdonald  symmetric function $P_\lambda(x;q,t)$ is uniquely characterized by 
the conditions \cite[Chap. VI, (4.7)]{M:1995}.
\begin{align*}
& P_\lambda= m_\lambda+\sum_{\mu<\lambda} u_{\lambda\mu}m_\mu
\qquad (u_{\lambda\mu}\in \bbF),\\
& \langle P_\lambda,P_\mu \rangle_{q,t}=0\qquad (\lambda\neq \mu). 
\end{align*}
\end{fct}

Denote 
$Q_\lambda \seteq P_\lambda / \langle P_\lambda,P_\lambda \rangle_{q,t}$.
Then $(Q_\lambda)$ and $(P_\lambda)$ are dual bases of $\Lambda_{\bbF}$.

The integral form $J_\lambda$ is defined 
by \cite[Chap. VI, (8.1),(8.$1'$),(8.3)]{M:1995}.
\begin{align}\label{eq:level1:J}
\nonumber
&J_\lambda \seteq c_\lambda P_\lambda=c'_\lambda Q_\lambda,\quad
\\
&c_\lambda \seteq
 \prod_{\square\in \lambda }
 (1-q^{a_\lambda(\square)}t^{\ell_\lambda(\square)+1}),\quad
c'_\lambda \seteq
 \prod_{\square\in \lambda }
 (1-q^{a_\lambda(\square)+1}t^{\ell_\lambda(\square)}).
\end{align}

As for the norms of $P_\lambda$ and $J_\lambda$, we have
 \cite[Chap. VI, (6.19)]{M:1995}
\begin{align}\label{eq:level1:norm}
\langle P_\lambda,P_\lambda\rangle_{q,t}= c'_\lambda / c_\lambda,\quad
\langle J_\lambda,J_\lambda\rangle_{q,t}= c'_\lambda  c_\lambda.
\end{align}

%%%%%%%%%%%%%%%%%%%%%%%%%%
\subsection{Level one representation of $\calU$}
\label{level_one_Fock}
Recall
the level one representation constructed over the space of Macdonald polynomials  
\cite{FHHSY:2009}.
Set $\FF\seteq\bbQ(q^{1/4},t^{1/4})$.

Let $\calH$ be the Heisenberg algebra over $\FF$ 
with generators $\{a_n \mid n\in\bbZ\}$ satisfying 
\begin{align*}
 [a_m,a_n]=m\dfrac{1-q^{|m|}}{1-t^{|m|}}\delta_{m+n,0} \, a_0.
\end{align*}
Let  $|0\rangle$ be the vacuum state
satisfying the annihilation conditions for 
the positive Fourier modes $a_n|0\rangle=0$ ($n\in\bbZ_{>0}$).
For a partition $\lambda=(\lambda_1,\lambda_2,\ldots)$, we 
denote 
$\ket{a_{\lambda} }=a_{-\lambda_1} a_{-\lambda_2}\cdots |0\rangle$ for short.
Denote by $\calF$ the Fock space 
having the basis $(\ket{a_{\lambda} })$.

As graded vector spaces, the space of the symmetric functions $\Lambda_{\FF}$
and the Fock space $\calF$ are isomorphic. 
We denote the isomorphism by $\iota$.
It is defined by
\begin{align}\label{eq:intro:iota}
\iota: \calF \simto \Lambda_{\FF},\quad 
      \ket{ a_{\lambda}} \mapsto p_\lambda.
\end{align}
We give an $\calH$-module structure on $\Lambda_{\FF}$ by setting
$a_0v=v$ and
\begin{align*}
a_{-n} v=p_n v,\quad
a_{n}  v=n\dfrac{1-q^n}{1-t^n}\dfrac{\partial v}{\partial p_n},\qquad (n>0,v\in \Lambda_{\FF}).
\end{align*}
In what follows we identify $\calF$ and $\Lambda_{\FF}$ 
as $\calH$ module
via $\iota$.

\begin{fct}[{\cite[Prop.~A.6]{FHHSY:2009}}]\label{fct:intro:level-one}
Set
\begin{align*}
&\eta(z)\seteq
\exp\Big( \sum_{n=1}^{\infty} \dfrac{1-t^{-n}}{n}a_{-n} z^{n} \Big)
\exp\Big(-\sum_{n=1}^{\infty} \dfrac{1-t^{n} }{n}a_n    z^{-n}\Big),\\
&\xi(z)\seteq
\exp\Big(-\sum_{n=1}^{\infty} \dfrac{1-t^{-n}}{n}(t/q)^{n/2}a_{-n} z^{n}\Big)
\exp\Big( \sum_{n=1}^{\infty} \dfrac{1-t^{n}}{n} (t/q)^{n/2} a_n z^{-n}\Big),\\
&\varphi^{+}(z)\seteq
\exp\Big(
 -\sum_{n=1}^{\infty} \dfrac{1-t^{n}}{n} (1-t^n q^{-n})(t/q)^{-n/4} a_n z^{-n}
    \Big),
\\
&\varphi^{-}(z)\seteq
\exp\Big(
 \sum_{n=1}^{\infty} \dfrac{1-t^{-n}}{n} (1-t^n q^{-n})(t/q)^{-n/4} a_{-n}z^{n}
    \Big).
\end{align*}
Let  $u\in {\widetilde{\bbF}}^*$.  We have a level one
representation $\rho_u(\cdot)$ of $\calU$ on $\calF$ by setting
\begin{align*}
&\rho_u(\gamma^{\pm 1/2})=(t/q)^{\pm 1/4},\quad 
 \rho_u(\psi^\pm(z))=\varphi^\pm(z), 
\quad
%\\
%&
\rho_u(x^+(z))=u\, \eta(z),\quad 
 \rho_u(x^-(z))=u^{-1} \xi(z).
\end{align*}
We denote this left $\calU$-module by $\calF_u$.
\end{fct}

\begin{fct}[{\cite{AMOS:1995}\cite{S:2006}}] 
The $x_0^+$ 
is identified with the first-order Macdonald difference operator 
(under the isomorphism $\iota:\calF_u\simto\Lambda_{\FF}
:|P_\lambda\rangle \mapsto P_\lambda$, see  \eqref{eq:intro:iota})
\begin{align}\label{eq:x0_eigen}
x_0^+ |P_\lambda\rangle
=u \ve_\lambda|P_\lambda\rangle,\quad
\ve_\lambda\seteq 1+(t-1) \sum_{i=1}^{\ell(\lambda)} (q^{\lambda_i}-1)t^{-i}.
\end{align}

\end{fct}

The dual Fock space $\calF^*$ is 
defined in a similar manner. Let $\langle 0|$ be the dual vacuum state
satisfying the annihilation conditions for 
the negative Fourier modes $\langle 0|a_n=0$ ($n \in \bbZ_{<0}$).
For a partition $\lambda=(\lambda_1,\lambda_2,\cdots)$, 
write  $\bra{a_\lambda}=\langle 0|\cdots a_{\lambda_2}a_{\lambda_1}$ 
for short. The $(\bra{a_\lambda})$ is a basis of $\calF^*$.
By the homomorphism $\rho_u$,
$\calF^*$ becomes a right $\calU$-module.

We have the compatibility between the 
Macdonald scalar product and the Fock pairing:
\begin{align*}
\langle p_\lambda,p_\mu \rangle_{q,t}=
\langle a_\lambda| a_{\mu} \rangle.
\end{align*}

%%%%%%%%%%%%%%%%%%%%%%%%%%%
\subsection{Integral basis $\ket{K_\lambda}$ for the level one case}
\label{K_lambda:level_one}

One of our motivations of this paper is to study the 
integral form $J_\lambda=c_\lambda P_\lambda$ 
of the Macdonald symmetric function,
and its higher level analogues, from the point of view of the Ding-Iohara algebra $\calU$. 
A point is how one can understand the mysterious
normalization of $J_\lambda$.

The standard normalization of the Macdonald symmetric function
is based on the lower triangular expansion 
$P_\lambda=m_\lambda+\sum_{\mu<\lambda} u_{\lambda\mu} m_\mu$ 
with respect to the dominance ordering.
Set the integral form by $J_\lambda=c_\lambda P_\lambda$, then 
the scalar product $\langle J_\lambda,J_\lambda\rangle_{q,t}=
c'_\lambda c_\lambda$ is a polynomial in $q$ and $t$. 
As we will observe shortly, we have a similar polynomiality 
in all the matrix elements of our vertex operator
with respect to the integral forms. 
At first glance, 
it seems that we need to face the 
problem of understanding the $c_\lambda$ from the algebra $\calU$.
We, however, bypass it by introducing 
a Poincar\'e-Birkhoff-Witt-type basis for $\calF_u$.
\medskip

For simplicity of display, we treat separately the level one case here. 
We omit writing the symbol $\rho_u$ from our formulas.
For any partition $\lambda$, set $|X_{\lambda}\rangle$ by 
\begin{align*}
|X_{\lambda}\rangle=
x_{-\lambda_1}^+x_{-\lambda_2}^+\cdots x_{-\lambda_{\ell(\lambda)}}^+
|0\rangle.
\end{align*}
For the dual space, we set
\begin{align*}
\bra{X_{\lambda}}=
\bra{0}x_{\lambda_{\ell(\lambda)}}^+\cdots x_{\lambda_2}^+x_{\lambda_1}^+.
\end{align*}

\begin{prp}
The $(|X_{\lambda}\rangle)$ 
(resp. $(\bra{X_{\lambda}})$) is a basis of $\calF$ (resp. $\calF^*$). 
\end{prp}

On $\calF_u$, we can expand the eigenfunctions of the operator $x^+_0$,
namely the $\ket{P_\lambda}$'s, 
with respect to the basis $(|X_{\lambda}\rangle)$. 
Set
\begin{align*}
\ket{K_\lambda} =\ket{X_{(1^{|\lambda|})}} +
\sum_{\mu>(1^{|\lambda|})}
c_{\lambda\mu}(u)\ket{X_\mu},\qquad
x^+_0|K_\lambda\rangle= u \ve_\lambda |K_\lambda\rangle ,
\end{align*}
were $c_{\lambda\mu}(u)\in \bbF[u]$.
Namely, we normalize the eigenfunctions $\ket{K_\lambda}$ in such a way that 
the coefficient of $\ket{X_{(1^{|\lambda|})}} $ is one.

Similarly on the dual space $\calF^{*}_{u}$, set
\begin{align*}
\langle K_\lambda| =
\bra{X_{(1^{|\lambda|})}} +
\sum_{\mu>(1^{|\lambda|})}
c_{\lambda\mu}(u)\bra{X_\mu},\qquad
\langle K_\lambda|x^+_0= u \ve_\lambda \langle K_\lambda|.
\end{align*}

\begin{eg}
For $|\lambda|\leq 2$, we have
\begin{align*}
&\ket{K_{(1)}}=\ket{X_{(1)}}=-t^{-1} u \ket{J_{(1)}},\\
&\ket{K_{(2)}}=\ket{X_{(1^2)}}+
{(q-1)u\over t}\ket{X_{(2)}}
=t^{-2} u^2 \ket{J_{(2)}},\\
&\ket{K_{(1^2)}}=\ket{X_{(1^2)}}+
{q(t-1)u\over t}\ket{X_{(2)}}
=t^{-3} u^2 \ket{J_{(1^2)}}.
\end{align*}
\end{eg}

In this paper we use the following 
notation for the so-called `Nekrasov factor.'
\begin{dfn}\label{dfn:N}
For a pair of partitions $(\lambda,\mu)$ and an indeterminate $u$, set
\begin{align*}
N_{\lambda,\mu}(u)
&\seteq
\prod_{(i,j)\in \lambda}(1-u q^{-\mu_i+j-1}t^{-\lambda'_j+i}) \cdot 
\prod_{(k,l)\in \mu}(1-u q^{\lambda_k-l}t^{\mu'_l-k+1}) \\
&=
\prod_{\square\in \lambda}
(1-u q^{-a_\mu(\square)-1}t^{-\ell_\lambda(\square)}) \cdot 
\prod_{\blacksquare \in \mu}
(1-u q^{a_\lambda(\blacksquare)}t^{\ell_\mu(\blacksquare)+1}).
\end{align*}
\end{dfn}

\begin{prp}\label{prop:level1:K}
We have
\begin{align*}
&\ket{K_\lambda} =
 (-u/t)^{|\lambda|} t^{-n(\lambda)} \ket{J_\lambda}, %\cdot 1,
\quad
 \bra{  K_\lambda} =
 (-u)^{|\lambda|} t^{-n(\lambda)} % 1\cdot 
\bra{ J_\lambda},
\\
&\langle K_\lambda| K_\lambda\rangle =
 (-u^2)^{|\lambda|} q^{n(\lambda')}t^{-n(\lambda)}N_{\lambda,\lambda}(q/t).
\end{align*}
\end{prp}

The proof is due to the specialization technique 
of \cite[Chap. VI, (6.17)]{M:1995}.
The detail will appear elsewhere.

%%%%%%%%%%%%%%%%%%%%%%%%%%%%%%%%%%%
\subsection{Vertex operator for the level one case}
\label{VO:level_one}
We state our definition of the level one vertex operator $\Phi_u^v(w)$ in terms of the
endomorphism $T(u,v)$.

\begin{dfn}\label{defVOlevelone}
Define the vertex operator $\Phi(w)$ by the conditions
\begin{align*}
&\Phi(w)=\Phi_u^v(w): 
 \calF_{u} \longrightarrow
 \calF_{v},
\\
&\Phi(w)\ket{0}=\ket{0}+O(w),
\\
&T(vw,q^{-1}t uw)(a)\Phi(w)
=\Phi(w)T(q^{-1}t vw,uw)(a)
\quad (\forall\, a\in \calU).
\end{align*}
\end{dfn}

In terms of $\eta(z),\xi(z),\varphi^\pm(z)$, the permutation relations are explicitly written as
\begin{align*}
&(1-vw/z)v\eta(z)\Phi(w)=(1-q^{-1}tvw/z)\Phi(w)u\eta(z),\\
&(1-(t/q)^{3/2}uw/z)v^{-1}\xi(z)\Phi(w)=(1-(t/q)^{1/2}uw/z)\Phi(w)u^{-1}\xi(z),
\\
&(1-(t/q)^{-1/4}vw/z)(1-(t/q)^{7/4}uw/z)\varphi^+(z)\Phi(w)\\
&\qquad=
(1-(t/q)^{3/4}vw/z)(1-(t/q)^{3/4}uw/z)\Phi(w)\varphi^+(z),
\\
&(1-(t/q)^{1/4}vw/z)(1-(t/q)^{5/4}uw/z)\varphi^-(z)\Phi(w)\\
&\qquad=
(1-(t/q)^{5/4}vw/z)(1-(t/q)^{1/4}uw/z)\Phi(w)\varphi^-(z).
\end{align*}
{}From these, one immediately finds that 
the $\Phi(w)$ can be uniquely expressed 
in terms of a normal ordered exponent of the Heisenberg generators.

\begin{prp}\label{Phi=}
We have
\begin{align}\label{eq:level1:Phi}
&\Phi(w) =
 \exp \Bigl(
 -\sum_{n=1}^{\infty} \dfrac{1}{n}\dfrac{v^n-(t/q)^{n}u^n}{1-q^n} a_{-n}w^n 
      \Bigr)
 \exp\Bigl(
  \sum_{n=1}^{\infty} \dfrac{1}{n}\dfrac{v^{-n}- u^{-n}}{1-q^{-n}} a_{n}w^{-n}
     \Bigr).
\end{align}
\end{prp}

Now we are ready to state our main result.
\begin{prp}\label{conj:level1}
Let $J_\lambda$ be the integral form of the Macdonald polynomial.
Then we have
\begin{align*}
\bra{ J_\lambda } \Phi(w) \ket{J_\mu} =
N_{\lambda,\mu}(qv/tu)
w^{|\lambda|-|\mu|}(tu/q)^{|\lambda|}(-v/q)^{-|\mu|} t^{n(\lambda)}q^{n(\mu')}.
\end{align*}
\end{prp}

This Proposition and Proposition \ref{prop:level1:K} give us
\begin{align*}
\bra{  K_\lambda} \Phi(w) \ket{K_\mu} =
N_{\lambda,\mu}(q v/t u)
(-t u v w/q)^{|\lambda|}(t v w/q)^{-|\mu|} 
u^{|\mu|} t^{-n(\mu)}q^{n(\mu')}.
\end{align*}

\begin{rmk}
%(1)
Proposition \ref{conj:level1} 
is nothing but the $K$-theoretic analogue of \cite{CO:2008}.
In fact, one can prove this based on  their argument and 
the geometric realization of Ding-Iohara algebra 
on $\oplus_n K^{\bbT}(\Hilb_n(\bbC^2))$.
The $\Phi(w)$ is essentially the same with 
the operator constructed from certain virtual bundle in \cite{SV:2009}.
The proof will appear elsewhere.
\end{rmk}

Consider the composition of the vertex operators
\begin{align*}
&\Phi_v^w(z_1)\Phi_u^v(z_2): \calF_{u}\longrightarrow \calF_{v}
\longrightarrow \calF_{w}.\qquad
\end{align*}
We have from Proposition \ref{conj:level1} 
\begin{align} 
\bra{0} \Phi_v^w(z_1)\Phi_u^v(z_2)\ket{0}
=
\sum_{\lambda} 
{N_{\emptyset,\lambda}(qw/tv)N_{\lambda,\emptyset}(qv/tu)\over 
N_{\lambda,\lambda}(q/t)} 
(uz_2/wz_1)^{|\lambda|}.\label{2pt}
\end{align}
The right hand side of (\ref{2pt}) coincides with the instanton part of the 
5D $\U(1)$ Nekrasov partition function with $N_f=2$ fundamental matters
(see \cite[\S 5]{AY:2010:2}). 
See Remark \ref{5D} below as for the higher level case.

%%%%%%%%%%%%%%%%%%%%%
\subsection{Examples of the calculation of the matrix elements of $\Phi(w)$}
We show some examples of calculating the 
matrix elements of  $\Phi(w)$.

On $\calF_u$, we have
\begin{align}
&\sum_{l\geq 0}f_l x^+_{m-l} x^+_{n+l}=
\sum_{l\geq 0}f_l x^+_{n-l} x^+_{m+l},\label{fxx}\\
&f_0=1,\qquad f_l={(1-q)(1-t^{-1})(1-q^lt^{-l} )\over 1-q t^{-1}}\quad 
\mbox{for } l=1,2,3,\ldots.
\nonumber
\end{align}
The permutation rule for $x^{+}_n$ and $\Phi(w)=\Phi_u^v(w)$ reads
\begin{align}\label{permu}
(x^{+}_n-vw x^+_{n-1})\Phi(w)=\Phi(w)(x^{+}_n-q^{-1}tvw x^+_{n-1}). 
\end{align}
We have $\bra{0}\Phi(w)\ket{0}=1$, $x^+_0 \ket{0}=u\ket{0}$, 
$x^+_{n} \ket{0}=0$ ($n=1,2,\ldots$), 
and 
$\bra{0}x^+_0 =v\bra{0}$, $\bra{0}x^+_{-n} =0$ 
($n=1,2,\ldots$).

{}From (\ref{permu}) written for $n=1$, we have
$\bra{0}(x^{+}_1-vw x^+_{0})\Phi(w)\ket{0}=
\bra{0}\Phi(w)(x^{+}_1-q^{-1}tvw x^+_{0})\ket{0}$. Hence we have
\begin{align*}
\bra{X_{(1)}}\Phi(w)\ket{X_\emptyset} =\bra{0}x^+_{1}\Phi(w)\ket{0}=
vw(v-q^{-1}t u).
\end{align*}

{}From (\ref{fxx}) written for $m=1,n=0$ and  $m=1,n=-1$, we have
$\bra{0}x^+_1x^{+}_0=v(1-f_1) \bra{0}x^+_1 $
and 
$\bra{0}x^+_1x^{+}_{-1}=- v^2f_1 \bra{0} $.
Then, from (\ref{permu}) written for $n=0$, we have
$\bra{0}x^+_1(x^{+}_0-vw x^+_{-1})\Phi(w)\ket{0}=
\bra{0}x^+_1\Phi(w)(x^{+}_0-q^{-1}tvw x^+_{-1})\ket{0}$. Hence we have
\begin{align*}
&v (1-f_1)\bra{0}x^+_{1}\Phi(w)\ket{0}+
v^3w f_1\bra{0}\Phi(w)\ket{0}\\
&=
u \bra{0}x^+_{1}\Phi(w)\ket{0}
-q^{-1}tvw  \bra{0}x^+_{1}\Phi(w)x^+_{-1}\ket{0},
\end{align*}
namely
\begin{align*}
\bra{X_{(1)}}\Phi(w)\ket{X_{(1)}} = \bra{0}x^+_{1}\Phi(w)x^+_{-1}\ket{0}=-u^2
(1-q v/u)(1-v/tu).
\end{align*}

%%%%%%%%%%%%%%%%%%%%%%%%%%%%%%
\section{Level $m$ representation}\label{subsubsec:tensored}
One can easily guess what should be the
higher level counterparts of the intertwining properties in Definition \ref{defVOlevelone},
and the integral basis $\ket{K_\lambda}$. 
By some brute force computations, we observed that 
the AGT phenomena may exist also for the higher level cases,
namely, all the matrix elements 
of the vertex operator with respect to $\ket{K_\lambda}$
are factorized and written 
in terms of the function $N_{\lambda,\mu}(u)$.  

%%%%%%%%%%%%%%%%%%%%%%%%%%%%%%%%%
\subsection{`PBW-type basis' for the level $m$ case}

Let $m$ be a positive integer 
and  $\bfu=(u_1,u_2,\ldots,u_m)$ be an $m$-tuple of parameters.
Consider the $m$-fold tensor representation 
$\rho_{u_1}\otimes\rho_{u_2}\otimes\cdots \otimes\rho_{u_m}$ 
on $\calF^{\otimes m}$.
Define $\Delta^{(m)}$ inductively by
$\Delta^{(1)}\seteq \id$, $\Delta^{(2)}\seteq \Delta$ and
$\Delta^{(m)}\seteq(\id \otimes \cdots 
\otimes{\rm id}\otimes \Delta)\circ \Delta^{(m-1)}$.
\begin{dfn}\label{dfn:level_m_rep}
Define the morhpism $\rho_{\bfu}^{(m)}$ by
\begin{align*}
\rho_{\bfu}^{(m)}\seteq
(\rho_{u_1}\otimes\rho_{u_2}\otimes\cdots \otimes\rho_{u_m})
\circ\Delta^{(m)}.
\end{align*}
We denote by $\calF_{\bfu}$ (resp. $\calF_{\bfu}^*$)
the left (resp. right) $\calU$-module 
on $\calF^{\otimes m}$ (resp. ${\calF^*}^{\otimes m}$)
given by $\rho_{\bfu}^{(m)}$.
These representations are of level $m$,
and we call them \emph{the} level $m$ representations. 
\end{dfn}

Set
\begin{align*}
X^{(1)}(z)\seteq
\rho_{\bfu}^{(m)}(x^+(z))
=(\rho_{u_1}\otimes\rho_{u_2}\otimes\cdots \otimes\rho_{u_m})
\circ\Delta^{(m)}(x^+(z)).
\end{align*}
Then we have 
\begin{align}\label{eq:whit:X1}
X^{(1)}(z)=\sum_{i=1}^m u_i \widetilde{\Lambda}_i(z),
\end{align} 
where
\begin{align}\label{eq:whit:Lambda}
\widetilde{\Lambda}_i(z)
&\seteq
 \varphi^-(p^{-1/4}z)\otimes\varphi^-(p^{-3/4}z)\otimes\cdots
 \otimes \varphi^-(p^{-(2i-3)/4}z)
 \otimes \eta(p^{-(i-1)/2}z)\otimes 1\otimes \cdots\otimes 1.
\end{align}
Here $p\seteq q/t$ 
and $\eta(p^{-(i-1)/2}z)$ sits in the $i$-th tensor component.
(See \cite[Lemma 2.6]{FHSSY:2010}.)
For $k=2,3,\ldots$, set further
\begin{align*}
X^{(k)}(z)\seteq
X^{(1)}(p^{k-1}z)\cdots X^{(1)}(pz)X^{(1)}(z).
\end{align*}
Then  for $k=1,2,\ldots, m$ we have 
\begin{align*}
X^{(k)}(z)=\sum_{1\le i_1<i_2<\cdots<i_k\le m}
 u_{i_1}u_{i_2}\cdots u_{i_k}
 :\widetilde{\Lambda}_{i_1}(z)
 \widetilde{\Lambda}_{i_2}(p z)
 \cdots
 \widetilde{\Lambda}_{i_k}(p^{k-1} z):,
\end{align*}
and $0=X^{(m+1)}(z)=X^{(m+2)}(z)=\cdots$.
Here $:*:$ denotes the usual normal ordering in the Heisenberg algebra $\calH$.
Define the Fourier components $X^{(k)}_i$ of $X^{(k)}(z)$ by
\begin{align*}
X^{(k)}(z)=\sum_{i\in\bbZ} X^{(k)}_i z^{-i}.
\end{align*}

\begin{rmk}
As for the connection between the $X^{(i)}(z)$'s and
the deformed $\calW_m$ generators, see \cite{FHSSY:2010}.
\end{rmk}

\begin{dfn}
Let $\vlambda=(\lambda^{(1)},\lambda^{(2)},\ldots,\lambda^{(m)})$
be an $m$-tuple of partitions with 
$\lambda^{(k)}=(\lambda^{(k)}_1,\lambda^{(k)}_2,\ldots)$.
We set
\begin{align*}
&\ket{X_{\Vec{\lambda}}}\seteq
X^{(1)}_{-\lambda^{(1)}_1}X^{(1)}_{-\lambda^{(1)}_2}\cdots
X^{(2)}_{-\lambda^{(2)}_1}X^{(2)}_{-\lambda^{(2)}_2}\cdots
X^{(m)}_{-\lambda^{(m)}_1}X^{(m)}_{-\lambda^{(m)}_2}\cdots
\ket{\bf 0},\\
&\bra{X_{\Vec{\lambda}}}\seteq
(q/t)^{ \sum_{k=1}^m (k-1) | \lambda^{(k)} |}%(q/t)^{\sum_{i=2}^m i|\lambda^{(i)}|}
\bra{\bf 0}
\cdots X^{(m)}_{\lambda^{(m)}_2}X^{(m)}_{\lambda^{(m)}_1}
\cdots X^{(2)}_{\lambda^{(2)}_2}X^{(2)}_{\lambda^{(2)}_1}
\cdots X^{(1)}_{\lambda^{(1)}_2}X^{(1)}_{\lambda^{(1)}_1},
\end{align*}
where $\ket{\bf 0}\seteq \ket{0}^{\otimes m}$ and 
$\bra{\bf 0}\seteq \bra{0}^{\otimes m}$.
\end{dfn}
  
\begin{con}
The $(\ket{X_{\Vec{\lambda}}})$ (resp. $(\bra{X_{\Vec{\lambda}}})$) is a basis of
$\calF_{\bf u}$ (resp. $\calF_{\bf u}^*$).
\end{con}

%%%%%%%%%%%%%%%%%%%%%%%%%%%%
\subsection{Partial orderings}
\label{Partial_Ordering}
As in the case of level one representation, we study the 
eigenfunctions of the operator $X^{(1)}_0=\rho_{\bfu}^{(m)}(x^{+}_{0})$ 
on the spaces $\calF_{\bf u}$ 
and  $\calF^*_{\bf u}$ .
A remark is in order.
We can not regard the $X^{(1)}_0$ as a self adjoint operator,
because of the structure of the coproduct.
Hence we need to consider
the left eigenfunctions in $\calF_{\bf u}$ and 
the right eigenfunctions in $\calF_{\bf u}^*$ separately.

For $\vlambda=(\lambda^{(1)},\lambda^{(2)},\ldots,\lambda^{(m)})$ with 
$\lambda^{(k)}=(\lambda^{(k)}_1,\lambda^{(k)}_2,\ldots)$,
we denote the total number of boxes by
$|\vlambda| \seteq \sum_{k=1}^m |\lambda^{(k)}|$.

\begin{dfn}
Introduce two partial orderings $\geq^R$ and $\geq^L$ 
on the $m$-tuples of partitions by 
\begin{align}\label{eq:levelm:order}
\begin{split}
&\vlambda \ge^R \vmu 
\stackrel{\text{def}}{\Longleftrightarrow} 
|\vlambda|=|\vmu| \text{ and }  
\\
&
\qquad\qquad \quad|\lambda^{(1)}|+\cdots+|\lambda^{(j-1)}|+\sum_{k=1}^{i} \lambda_k^{(j)}
\ge
|\mu^{(1)}|+\cdots+|\mu^{(j-1)}|+\sum_{k=1}^{i} \mu_k^{(j)}
\\
& \qquad\qquad\quad\text{for all } i \ge1, 1\le j\le m,\\
&\vlambda \ge^L \vmu 
\stackrel{\text{def}}{\Longleftrightarrow} 
|\vlambda|=|\vmu| \text{ and }  
\\
&
\qquad\qquad \quad|\lambda^{(m)}|+\cdots+|\lambda^{(j+1)}|+\sum_{k=1}^{i} \lambda_k^{(j)}
\ge
|\mu^{(m)}|+\cdots+|\mu^{(j+1)}|+\sum_{k=1}^{i} \mu_k^{(j)}
\\
& \qquad\qquad\quad\text{for all } i \ge1, 1\le j\le m.
\end{split}
\end{align}
\end{dfn}

\begin{eg}
We consider the case $>^L$ and  denote it by $>$ for short.
In the case $m=2$ and $|\vlambda|\leq 3$, we have
\begin{align*}
&(\emptyset,(1)) > ((1),\emptyset),
\\
& (\emptyset,(2)) > (\emptyset,(1^2))> ((1),(1))
 >((2),\emptyset) > ((1^2),\emptyset), 
\\
&(\emptyset,(3)) > (\emptyset,(21)) 
\genfrac{}{}{0pt}{1}{\displaystyle > ((1),(2)) >}{\displaystyle > (\emptyset,(1^3))>}
((1),(1^2)) > ((2),(1)) 
\genfrac{}{}{0pt}{1}{\displaystyle >((3),(\emptyset))>}{\displaystyle > ((1^2),(1)) > }
((21),\emptyset)>((1^3),\emptyset).
\end{align*}
\end{eg}

%%%%%%%%%%%%%%%%%%%%%
\subsection{Eigenfunctions}
\label{Mac_level_m}

For an $m$-tuple of partitions $\vlambda$, set
\begin{align*}
 m_{\Vec{\lambda}}=
 m_{\lambda^{(1)}}\otimes m_{\lambda^{(2)}}\otimes \cdots 
                  \otimes m_{\lambda^{(m)}} \in \Lambda^{\otimes m},
\end{align*}
where $m_{\lambda^{(i)}}$'s are the monomial symmetric functions.
Via the isomorphism $\iota^{\otimes m}$ (see \eqref{eq:intro:iota}), 
we identify 
 $ m_{\Vec{\lambda}}\in \Lambda^{\otimes m}$
with the corresponding vector $\ket{m_{\Vec{\lambda}}}\in {\calF}_{\bf u}$
or $\bra{m_{\Vec{\lambda}}}\in {\calF}_{\bf u}^*$.

 \begin{prp}\label{expan}
We have
\begin{align*}
&X_0^{(1)}\ket{m_{\Vec{\lambda}}}
=  \sum_{\Vec{\mu}\leq^L \Vec{\lambda}} \alpha_{\Vec{\lambda}\Vec{\mu}}({\bf u})
  \ket{ m_{\Vec{\mu}}},\\
&\bra{m_{\Vec{\lambda}}}X_0^{(1)}
=  \sum_{\Vec{\mu}\leq^R \Vec{\lambda}} \beta_{\Vec{\lambda}\Vec{\mu}}({\bf u})
  \bra{ m_{\Vec{\mu}}},
\end{align*}
for some  $\alpha_{\Vec{\lambda}\Vec{\mu}}({\bf u}),\beta_{\Vec{\lambda}\Vec{\mu}}({\bf u})
\in \FF[u_1,u_2,\ldots,u_m]$.
\end{prp}

\begin{prp}
(1) For any $m$-tuples of partitions $\vlambda$, 
a vector $\ket{P_{\Vec{\lambda}}}\in\calF_{\bfu}$ is uniquely characterized by 
\begin{align*}
&\ket{P_{\Vec{\lambda}}}
= \ket{ m_{\Vec{\lambda}}}+
                  \sum_{\Vec{\mu}<^L\Vec{\lambda}} a_{\Vec{\lambda}\Vec{\mu}}({\bf u})
    \ket{ m_{\Vec{\mu}}},
     \qquad (a_{\Vec{\lambda}\Vec{\mu}}({\bf u})\in \FF(u_1,u_2,\ldots,u_m)),\\
&X^{(1)}_0 \ket{ P_{\Vec{\lambda}}}
=\ve_{\vlambda,\bfu}\ket{P_{\Vec{\lambda}}},
 \qquad
 \ve_{\Vec{\lambda},\bfu}\seteq
 \sum_{k=1}^m u_k \ve_{\lambda^{(k)}}.
\end{align*}
(2)
 For any $m$-tuples of partitions $\vlambda$, 
a vector $\bra{P_{\Vec{\lambda}}}\in\calF_{\bfu}^*$ is uniquely characterized by 
\begin{align*}
&\bra{P_{\Vec{\lambda}}}
= \bra{m_{\Vec{\lambda}}}+
                  \sum_{\Vec{\mu}<^R\Vec{\lambda}} b_{\Vec{\lambda}\Vec{\mu}}({\bf u})
    \bra{ m_{\Vec{\mu}}},
     \qquad (b_{\Vec{\lambda}\Vec{\mu}}({\bf u})\in \FF(u_1,u_2,\ldots,u_m)),\\
&\bra{ P_{\Vec{\lambda}}}X^{(1)}_0 
=\ve_{\vlambda,\bfu}\bra{P_{\Vec{\lambda}}}.
\end{align*}
(3)
We have
\begin{align*}
\langle P_{\vlambda} | P_{\vmu}\rangle=
\prod_{k=1}^m \dfrac{c'_{\lambda^{(k)}}}{c_{\lambda^{(k)}}}
\cdot \delta_{\vlambda,\vmu}.
\end{align*}
\end{prp}

For (1) and (2), it is enough to prove Proposition \ref{expan}. 
The detail will appear elsewhere. Once we have (1) and (2), for the proof of (3) 
we only need to recall 
the norm of $P_\lambda$ \eqref{eq:level1:norm}.

The vector $\ket{ P_{\vlambda}}$ can be considered 
as a higher level analogue of the Macdonald symmetric function $P_\lambda$.

\begin{eg}
Consider the case $m=2$. 
We denote by $\ket{P_{\lambda_1}\otimes P_{\lambda_2}}$ 
the image of $P_{\lambda_1}\otimes P_{\lambda_2}$ 
in $\calF_{u_1}\otimes\calF_{u_2}$ under the isomorphism $\iota^{\otimes 2}$.
Below we give some examples of the vectors $\ket{P_{(\lambda_1,\lambda_2)}}$ 
expanded in terms of $(\ket{P_{\lambda_1}\otimes P_{\lambda_2}})$. 

First we trivially have $\ket{P_{(\emptyset,\emptyset)}}=\ket{1\otimes 1}$.
For $|\vlambda|=1$, we have
\begin{align*}
 \ket{P_{((1),\emptyset)}}
=&\ket{P_{(1)}\otimes 1},\\
 \ket{P_{(\emptyset,(1))}}
=&\ket{1\otimes P_{(1)} }
  +(q/t)^{1/2}\dfrac{(t-q)u_2}{q(u_1-u_2)}\ket{P_{(1)}\otimes 1}.
\end{align*}
For $|\vlambda|=2$, we have
\begin{align*}
 \ket{P_{((1^2),\emptyset)}}
=&\ket{P_{(1^2)} \otimes 1},\\
 \ket{P_{((2),\emptyset)}}
=&\ket{P_{(2)}\otimes 1},
\\
 \ket{P_{((1),(1))}} 
=&\ket{P_{(1)}\otimes P_{(1)}}
 +(q/t)^{1/2}\dfrac{(1-q)(t+1)t(t-q)u_2}{q(1-q t)(u_1-t u_2)} 
 \ket{ P_{(1^2)}\otimes 1}\\
 &+(q/t)^{1/2}\dfrac{(t-q)u_2}{q(q u_1-u_2)} %\times \sqrt{q/t}
\ket{  P_{(2)}\otimes 1},
\\
 \ket{P_{(\emptyset, (1^2))} }
=&\ket{1\otimes P_{(1^2)}}
 +(q/t)^{1/2}\dfrac{(t-q)u_2}{q(t u_1-u_2)}  %\times \sqrt{q/t}
 \ket{ P_{(1)}\otimes P_{(1)}}
\\
 &+\dfrac{(t-q)((q^2 t-q t-q+t^2)u_2-q t(t^2-1) u_1) u_2}
       {q t (u_1-u_2)(1- q t)(t u_1-u_2)} \ket{P_{(1^2)}\otimes 1}\\
 &-\dfrac{(t-q)u_2}{q(t u_1-u_2)} \ket{P_{(2)}\otimes 1},
\\
 \ket{P_{(\emptyset,(2))}}
=&\ket{1\otimes P_{(2)}}
  -(q/t)^{1/2}\dfrac{(t-1)(1+q)(t-q)u_2}{(1-q t)(u_1-q u_2)}
   \ket{P_{(1)}\otimes P_{(1)}}
\\
 &+\dfrac{(q-t)(q (q^2 t-q+q t-t^2)u_2+(1-q^2) t u_1)u_2}
        {q t(1-q t)(u_1-u_2)(q u_2-u_1)} \ket{P_{(2)}\otimes 1}
\\
 &-\dfrac{(t^2-1)(1-q^2)(t-q)u_2}{(q u_2-u_1)(1-q t)^2}
 \ket{ P_{(1^2)}\otimes 1}.
\end{align*} 
For the case $|\vlambda|=3$, 
the partial ordering $>^L$ is not a total ordering.
Here we give five examples 
for the sake of demonstration:
\begin{align*}
 \ket{P_{((1^3),\emptyset)}}
=&\ket{P_{(1^3)}\otimes 1},
\\
 \ket{P_{((2,1),\emptyset)}}
=&\ket{P_{(2,1)}\otimes 1},
\\
 \ket{P_{((3),\emptyset)}}
=&\ket{P_{(3)}\otimes 1},
\\
 \ket{P_{((1^2),(1))}}
=&\ket{P_{(1^2)}\otimes P_{(1)}}
  +(q/t)^{1/2}\dfrac{(t-q)u_2}{q(q u_1-u_2)} % \times \sqrt{q/t]
 \ket{ P_{(2,1)}\otimes 1}
\\
 &+(q/t)^{1/2}\dfrac{(1-q)(t-q)(1-t^3)t^2u_2}{q(1-q t^2)(1-t)(u_1-t^2 u_2)}
   \ket{P_{(1^3)}\otimes 1},
\\
 \ket{P_{((2),(1))}}
=&\ket{P_{(2)}\otimes P_{(1)}}
  +(q/t)^{1/2}\dfrac{(t-q)u_2}{q(q^2 u_1-u_2)} % \times \sqrt{q/t]
 \ket{ P_{(3)}\otimes 1}
\\
  &+(q/t)^{1/2}\dfrac{(1-q^2)(t-q)(1-q t^2)t u_2}{q(1-q t)(1-q^2 t)(u_1-t u_2)}
   \ket{P_{(2,1)}\otimes 1}.
\end{align*}
As for the dual eigenvectors, 
we have $\bra{P_{(\emptyset,\emptyset)}}=\bra{1\otimes 1}$, and
\begin{align*}
 \bra{P_{((1),\emptyset)}}
&=\bra{P_{(1)}\otimes 1}
  -(q/t)^{1/2}\dfrac{(t-q)u_2}{q(u_1-u_2)}\bra{1\otimes P_{(1)}},
\\
\bra{P_{(\emptyset,(1))}}
&=\bra{1\otimes P_{(1)} }
\end{align*}
for $|\vlambda|=1$. 
\end{eg}

%%%%%%%%%%%%%%%%%%%%%%%%%%%%%%%%%
\subsection{`Integral basis' $\ket{K_{\vlambda}}$ for the level $m$ case}
\label{K_lambda:level_m}

As in the level one case, we introduce the following 
normalization of the eigenvectors.
\begin{dfn}
Define the integral form $\ket{K_{\vlambda}}\in \calF_{\bfu}$ by 
\begin{align*}
X^{(1)}_0 \ket{K_{\vlambda}}=\ve_{\vlambda,\bfu}\ket{K_{\vlambda}},\qquad
\ket{K_{\Vec{\lambda}}}
=\bigl((X_{-1}^{(1)})^{|\vlambda|}+\cdots\bigr) \ket{\bf 0}.
\end{align*}
Similarly we define $\bra{K_{\vlambda}} \in \calF_{\bfu}^*$ by
\begin{align*}
\bra{K_{\vlambda}}X^{(1)}_0=\ve_{\vlambda,\bfu}\bra{K_{\vlambda}},\qquad
\bra{K_{\Vec{\lambda}}}
=\bra{\bf 0} \bigl((X_{1}^{(1)})^{|\vlambda|}+\cdots\bigr) .
\end{align*}
\end{dfn}

\begin{con}
We have
\begin{align}\label{eq:conj:norm:level:m}
\begin{split}
\langle K_{\Vec{\lambda}}|K_{\Vec{\lambda}}\rangle
\stackrel{?}{=}
&\bigl((-1)^m (t/q)^{m-1}e_m(\bfu) \bigr)^{|\Vec{\lambda}|}\\
&\times
\prod_{k=1}^m u_k^{-(m-2)|\lambda^{(k)}|}
q^{-(m-2)n(\lambda^{(k)'})}t^{(m-2)n(\lambda^{(k)})}
\times 
\prod_{i,j=1}^m
N_{\lambda^{(i)},\lambda^{(j)}}
(q u_i/tu_j).
\end{split}
\end{align}
with  $e_m(\bfu)\seteq u_1 u_2\cdots u_m$.
\end{con}

%%%%%%%%%%%%%%%%%%%%%
\subsection{Vertex operator for the level $m$ case}
\label{VO:level_m}
We extend the construction of the vertex operator $\Phi(w)$ for 
higher level cases.

\begin{dfn}\label{defVOlevelm}
Let ${\bf u}=(u_1,u_2,\cdots,u_m)$ and 
${\bf v}=(v_1,v_2,\cdots,v_m)$.
Define the vertex operator $\Phi(w)=\Phi_{\bf u}^{\bf v}(w)$ 
by
\begin{align*}
&\Phi(w): 
 \calF_{\bfu}=\calF_{u_1}\otimes\calF_{u_2}\otimes \cdots \otimes \calF_{u_m} 
 \longrightarrow
 \calF_{\bfv}=\calF_{v_1}\otimes\calF_{u_2}\otimes \cdots \otimes \calF_{v_m},
\\
&\Phi(w)\ket{\bf 0}
=\ket{\bf 0}+O(w),
\\
&T(e_m(\bfv)w,q^{-1}t e_m(\bfu)w)(a)\Phi(w)
=\Phi(w)T(q^{-1}t e_m(\bfv)w,e_m(\bfu)w)(a)
\quad (\forall\, a\in \calU).
\end{align*}
Here we used the symbols $e_m(\bfv)\seteq v_1 v_2 \cdots v_m$ 
and $e_m(\bfu)\seteq u_1 u_2\cdots u_m$.
\end{dfn}

Now we state our main conjecture.
\begin{con}\label{maincon}
(1) The $\Phi(w)$ exists uniquely.

(2)
We have the factorized matrix elements with respect to the integral forms as
\begin{align}\label{eq:conj:matrix:level:m}
\begin{split}
&\bra{ K_{\Vec{\lambda}}}\Phi(w)\ket{K_{\Vec{\mu}}} 
\stackrel{?}{=}
\bigl((-1)^m (t/q)^{m} e_m(\bfu) e_m(\bfv) w\bigr)^{|\Vec{\lambda}|}
 \bigl((t/q)  e_m(\bfv) w\bigr)^{-|\Vec{\mu}|}
\\
& %\times
%  \bigl((t/q)  e_m(\bfv) w\bigr)^{-|\Vec{\mu}|}
\qquad \qquad\qquad \qquad\times
  \prod_{k=1}^m v_k^{-(m-1)|\lambda^{(k)}|}
  u_k^{|\mu^{(k)}|}
  q^{-(m-1)n(\lambda^{(k)'})+n(\mu^{(k)'})}
  t^{(m-1)n(\lambda^{(k)})-n(\mu^{(k)})}
\\
&\qquad \qquad\qquad\qquad \times 
  \prod_{i,j=1}^m
  N_{\lambda^{(i)},\mu^{(j)}} (q v_i/tu_j).
\end{split}
\end{align}
\end{con}

Conjectures \eqref{eq:conj:matrix:level:m} and \eqref{eq:conj:norm:level:m} imply
\begin{align}
\bra{\bf 0} \Phi^{\mathbf{w}}_{\bfv}(z_2) \Phi^{\bfv}_{\bfu}(z_1) \ket{\bf 0}
&=
\sum_{\Vec{\lambda}}
 \dfrac{
  \bra{K_{\emptyset}}     \Phi^{\bfw}_{\bfv}(z_2) \ket{K_{\Vec{\lambda}}}
  \bra{K_{\Vec{\lambda}}} \Phi^{\bfv}_{\bfu}(z_1) \ket{K_{\emptyset}}}
  {\langle K_{\Vec{\lambda}} | K_{\Vec{\lambda}} \rangle} \label{4ptF}
\\
&\stackrel{?}{=}
\sum_{\Vec{\lambda}} 
 \biggr(\dfrac{e_m(\bfu) z_1}{e_m(\bfw) z_2}\biggr)^{|\Vec{\lambda}|}
 \prod_{i,j=1}^{m}
 \dfrac{
  N_{\emptyset,\lambda^{(j)}}(q w_i/t v_j)
  N_{\lambda^{(i)},\emptyset}(q v_i/t u_j)}
 {N_{\lambda^{(i)},\lambda^{(j)}}(q v_i/t v_j)} \nonumber
\\
&=
\sum_{\Vec{\lambda}} 
 \biggr(\dfrac{t^m}{q^m}\dfrac{e_m(\bfu) z_1}{e_m(\bfw) z_2}\biggr)^{|\Vec{\lambda}|}
 \prod_{i,j=1}^{m}
 \dfrac{
  N_{\emptyset,\lambda^{(j)}}(q w_i/t v_j)
  N_{\lambda^{(i)},\emptyset}(q v_i/t u_j)}
 {N_{\lambda^{(i)},\lambda^{(j)}}( v_i/ v_j)}. \nonumber
\end{align}

\begin{rmk}\label{5D}
The left hand side of (\ref{4ptF}) can be understood 
as a $q$-analogue of the four point correlation function of CFT.
The right hand side coincides with the instanton part of the 5D 
$\U(m)$ Nekrasov partition function with $N_f=2m$ fundamental matters 
(see \cite[\S 5]{AY:2010:2}).
Our main conjecture \ref{maincon} implies that 
we have a description of 
the five dimensional analogue of the AGT conjecture 
in terms of the level $m$ representation of the Ding-Iohara algebra.
\end{rmk}

\begin{rmk}
In \cite{AFLT:2011}, a good understanding is 
found about
the primary fields of the conformal field theory, the integrable structure, 
and the AGT conjecture. Their ideas and the main points are summarized as:
\begin{itemize}
\item to consider the extended algebra 
$\calA\seteq(\text{Virasoro algebra})\otimes\text{(Heisenberg algebra)}$, 
\item to study the integrable structure in $\calA$ and the complete eigenfunctions,
\item matrix elements of the primary field with respect to the eigen-basis,
\item factorization of the matrix elements in temrs of  the Nekrasov function.
\end{itemize} 
Note that on level two  ($m=2$),
$\calU$ is regarded as $\mbox{(deformed Virasoro)}\otimes \mbox{(Heisenberg algebra)}$.
Hence, it is expected that our level two case ($m=2$) 
can be regarded as a $q$-deformation of \cite{AFLT:2011}.
\end{rmk}

%%%%%%%%%%%%%%%%%%%%%%%%%%%%%%%%%%%%%%%%%%%%%%%%%%
%%%%%%%%%%%%%%%%%%%%%%%%%%%%%%%%%%%%%%%%%%%%%%%%%%

\section{Interlude: AGT conjecture}
\label{sec:intro:subsec:AGT}

\subsection{Four dimensional version}
In \cite{AGT:2010} a remarkable proposal, 
now called the AGT conjecture/relation, 
was given on the equivalence between the conformal block 
of the Liouville theory 
and the Nekrasov partition function.
Among the related investigations, 
Gaiotto proposed several degenerated versions in \cite{G:2009}.
Its simplest case  claims 
that the inner product $\langle G | G \rangle$ 
of a certain element $\ket{G}$ in the Verma module of Virasoro algebra 
coincides with the instanton part of the Nekrasov partition 
function $Z_{\text{pure} \SU(2)}^{\text{inst}}(\ve_1,\ve_2,\va;\Lambda)$
for the four dimensional $\mathcal{N}=2$ super-symmetric 
pure $\SU(2)$ gauge theory \cite{N:2003}.
%Actually, the element considered is a kind of Whittaker vector \cite{K:1978}
%in the Verma module of the Virasoro algebra.

\subsubsection{Whittaker vector for Virasoro algebra}
Recall the notion of the Whittaker vector for a finite dimensional 
Lie algebra $\frkg$. %, which was introduced in \cite{K:1978}. 
Let $\frkn$ be a maximal nilpotent Lie subalgebra of $\frkg$ 
and $\chi:\frkn\to\bbC$ be a character.
Let $V$ be any $U(\frkg)$-module.
Then a vector $w\in V$ is called a Whittaker vector with respect to $\chi$ 
if $x w=\chi(x)w$ for all $x\in\frkn$.

In \cite{G:2009}, analogue of Whittaker vectors was considered
for the Verma module of the Virasoro algebra.
Let $\Vir \seteq \bbC C \oplus \bigoplus_{n\in\bbZ}\bbC L_n$ 
be the Virasoro algebra with the relation
\begin{align*}
%&
[L_m,L_n]=(m-n)L_{m+n}+\dfrac{1}{12}(m^3-m)\delta_{m+n,0}\, C,
\quad
%&
[C,\Vir]=0.
\end{align*}
We have a triangular decomposition 
$\Vir=\Vir_{>0}\oplus\Vir_{0}\oplus\Vir_{<0}$
with $\Vir_{>0}\seteq \bigoplus_{n\in\bbZ_{>0}}\bbC L_n$,
     $\Vir_{0} \seteq \bbC C \oplus \bbC L_0$ 
and  $\Vir_{<0}\seteq \bigoplus_{n\in\bbZ_{<0}}\bbC L_n$.
The Verma module $M_{c,h}$ is a representation of $\Vir$ induced 
from $\bbC_{c,h}=\bbC\ket{c,h}$, the one dimensional representation 
of $\Vir_{>0}\oplus\Vir_{0}$ where $\Vir_{>0}$ acts trivially, 
$L_0$ acts by multiplication of $h$ 
and $C$ acts by multiplication of $c$.
 
Note that the elements $L_1,L_2\in\Vir_{>0}$ generate $\Vir_{>0}$.
Thus if we set $\frkn\seteq \Vir_{>0}$ in the above definition 
of the Whittaker vector,
then the homomorphism $\chi: \Vir_{>0}\to\bbC$ 
is determined by $\chi_1\seteq \chi(L_1)$ and $\chi_2\seteq \chi(L_2)$.
Then the Whittaker vector $v$ is an element of the completed
Verma module $\widehat{M}_{c,h}$ satisfying 
\begin{align*}
L_1 v=\chi_1 v,\quad L_2 v=\chi_2 v.
\end{align*}

The simplest case Gaiotto considered is the choice $\chi_2=0$, 
and we denote the corresponding Whittaker vector by $\ket{G}$.
Imposing a normalization condition and changing parameter $\chi_1$, 
we have
\begin{align}\label{eq:G:intro}
L_1 \ket{G}=\Lambda^2 \ket{G},\quad \ket{G}=\ket{c,h}+\cdots.
\end{align}
In fact such $\ket{G}$ is uniquely determined.

\subsubsection{Four dimensional Nekrasov partition function}
Recall \cite{N:2003} that Nekrasov's partition function 
$Z_{\text{pure} \SU(N)}^{\text{inst}}(\ve_1,\ve_2,\va;\Lambda)$
for four dimensional pure $\SU(N)$ gauge theory 
is defined to be the generating function of equivariant integrals 
over the $\SU(N)$ instanton moduli spaces $M_{N,n}$,
where $n$ is the instanton number:
\begin{align*}
Z_{\text{pure} \SU(N)}^{\text{inst}}(\ve_1,\ve_2,\va;\Lambda)
\seteq \sum_{n=0}^\infty \Lambda^{2 n N} \mbox{``} \int_{M_{N,n}}1\mbox{''}.
\end{align*}
Here the quoted integral is justified as follows 
(see also \cite{NY:2005:a}).
Let $M_0(N,n)$ be the Uhlenbeck partial 
compactification of 
the framed instanton moduli space of rank $N$ and instanton number $n$.
It is a (singular) affine variety and its complex dimension is $2 n N$.
It has an action of $\bbT\seteq(\bbC^{*})^{2}\times(\bbC^{*})^{N-1}$,
where  $(\bbC^{*})^2$ acts on $\bbC \bbP^2$
and $(\bbC^{*})^{N-1}$ acts on the framing.
The fixed point set $M_0(N,n)^{\bbT}$ consists of one point.
We denote by $\iota_{0*}: M_0(N,n)^{\bbT}\hookrightarrow M_0(N,n)$ 
the inclusion map.
Now consider the fundamental class 
$[M_0(N,n)]\in H_{4Nn}^{\bbT}(M_0(N,n),\bbC)$ 
of the moduli in the (Borel-Moore) equivariant homology group.
The inclusion map induces the pushforward 
$\iota_{0*}:H^{\bbT}_{*}(M_0(N,n)^{\bbT})\to H^{\bbT}_{*}(M_0(N,n))$.
By the localization theorem it is an isomorphism 
after tensoring the quotient field of $H_\bbT^*(\text{pt})$.
Now we define
\begin{align*}
Z_{\text{pure} \SU(N)}^{\text{inst}}(\ve_1,\ve_2,\va;\Lambda)
\seteq \sum_{n=0}^\infty \Lambda^{2 n N} \iota_{0*}^{-1}[M_0(N,n)].
\end{align*}
The obtained function can be considered as an element 
of the quotient field $H_\bbT^*(\text{pt})$.
We will write $H_\bbT^*(\text{pt})=\bbC[\ve_1,\ve_2,\va]$,
where $(\ve_1,\ve_2)$ corresponds to $(\bbC^{*})^2$ acting on $\bbC \bbP^2$,
and $\va$ corresponds to $(\bbC^{*})^{N-1}$ acting on the framing.
Thus $Z_{\text{pure} \SU(N)}^{\text{inst}}(\ve_1,\ve_2,\va;\Lambda)$ 
is an element of $\bbC(\ve_1,\ve_2,\va)[[\Lambda^{2N}]]$.

\subsubsection{AGT relation and its generalization}
The simplest AGT relation proposed in \cite{G:2009} is
\begin{align}\label{eq:AGT:pure4}
\langle G | G \rangle 
= Z_{\text{pure} \SU(2)}^{\text{inst}}(\ve_1,\ve_2,a;\Lambda),
\end{align}
where $\ket{G}$ is determined by \eqref{eq:G:intro},
and the parameters correspond as 
$c=13+6(\ve_1/\ve_2+\ve_2/\ve_1)$
and $h=\big((\ve_1+\ve_2)^2-a^2\big)/4 \ve_1 \ve_2$.
Several AGT relations including 
the above \eqref{eq:AGT:pure4} were proved 
by \cite{FL:2010} and \cite{HJS:2010}.
A special case of the original AGT conjecture for the conformal block 
was  proved in \cite{MMS:2011}.

The AGT conjecture implies an action of Virasoro algebra 
on the equivariant cohomology of the rank two instanton moduli.
The word `AGT conjecture/relation' means a conjectural existence 
of $\calW(p,\widehat{\frkg})$-algebra on the equivariant (intersection) 
cohomology on the moduli of parabolic ${}^{L}G$-sheaves.
Roughly speaking, this conjecture suggests a realization of `$\calW$-algebra' 
as the hidden symmetry of the `instanton moduli space'. 
See, for example, \cite{BFRF:2010}.

\subsection{Five dimensional version}
Let us mention another generalization of the AGT conjecture: 
$K$-theoretic analogue.
The paper \cite{AY:2010} proposed a conjecture which relates 
the instanton part of Nekrasov's five dimensional (or $K$-theoretic) 
pure $\SU(2)$ partition function 
$Z_{\text{pure} \SU(2)}^{\text{inst} K}(Q,q,t;\Lambda)$ 
(see \S \ref{sec:intro:subsec:AGT:subsubsec:Nek} below) 
to the deformed Virasoro algebra \cite{SKAO:1996}.
The conjecture claims that 
$Z_{\text{pure} \SU(2)}^{\text{inst} K}(Q,q,t;\Lambda)$ 
coincides with the inner product $\langle G;q,t | G;q,t \rangle$ 
of certain Whittaker vector $\ket{G;q,t}$ in the Verma module 
of the deformed Virasoro algebra. 
This state is a natural $q$-deformed analogue of \eqref{eq:G:intro}.

%Here we will recall the $K$-theoretic AGT conjecture in detail.
%Let us begin with the representation theoretic side.

\subsubsection{Recollection of the deformed Virasoro algebra}
First we introduce the deformed Virasoro algebra $\Vir_{q,t}$,
its Verma module $M_h$ and the Whittaker vector.

Let $q,t$ be two generic complex parameters.
Set $p\seteq q/t$ for simplicity. % and assume that $p$ is not a root of $-1$.
The deformed Virasoro algebra $\Vir_{q,t}$ 
 \cite{SKAO:1996} %and \cite{BP:1998})
is defined to be 
the associative $\bbC$-algebra generated by $\{T_n \mid n\in\bbZ\}$ 
and $1$ with relations 
\begin{align*}
[T_n,T_m]
=-\sum_{l=1}^\infty f_l(T_{n-l}T_{m+l}-T_{m-l}T_{n+l})
                   -\dfrac{(1-q)(1-t^{-1})}{1-p}(p^n-p^{-n})\delta_{m+n,0},
\end{align*}
where the coefficients $f_l$'s are determined 
by the following generating function:
\begin{align*}
\sum_{l=0}^\infty f_l x^l
=\exp\Big(\sum_{n=1}^\infty \dfrac{(1-q)(1-t^{-1})}{1+p^n}\dfrac{x^n}{n}\Big).
\end{align*}

Next we introduce a representation of this algebra.
For $h\in\bbC$, let $\ket{h}$ be a vector and define the action 
of $\Vir_{q,t}$ by  
%\begin{align*}
$1\ket{h}= \ket{h}$, 
$T_0\ket{h}= h\ket{h}$, %\quad
$T_n\ket{h}= 0$ ($\forall n\in\bbZ_{>0}$).
%\end{align*}
Then the Verma module $M_h$ for $\Vir_{q,t}$ is defined to be 
the $\Vir_{q,t}$-module generated by $\ket{h}$.

The dual (right) module $M_h^*$ is similarly defined.
It is generated by the highest weight vector $\bra{h}$ satisfying 
$\bra{h}1=\bra{h}$, $\bra{h}T_0=h\bra{h}$, 
and $\bra{h}T_n=0$ for any $n\in\bbZ_{<0}$.

Let us introduce the outer grading operator $d$ satisfying $[d,T_n]=-n T_n$.
Defining the action of $d$ on $M_h$ by $d \ket{h}=0$, 
we have the direct decomposition $M_h=\oplus_{n\in\bbZ_{\ge0}}M_{h,n}$ 
with respect to this grading.
$M_{h,n}$ has a basis consisting of 
$T_{-\lambda}\ket{h}
 \seteq T_{-\lambda_1}T_{-\lambda_2}\cdots T_{-\lambda_l}\ket{h}$
with $\lambda=(\lambda_1,\lambda_2,\ldots,\lambda_l)$, $|\lambda|=n$.
The dual representation $M_h^*$ also has a similar grading structure 
$M_h^*=\oplus_{n\in\bbZ_{\ge0}}M^*_{h,n}$,
and $M_h^*$ has a basis consisting of the vectors
$\bra{h}T_{\lambda}\seteq 
 \bra{h}T_{\lambda_l}\cdots T_{\lambda_2} T_{\lambda_1}$ 
indexed by partitions of $n$.

Let us denote by $\widehat{M}_h$ and $\widehat{M}^*_h$ 
the completions of $M_h$ and $M^*_h$ with respect to the grading above.
The deformed Gaiotto state $\ket{G;q,t}\in \widehat{M}_h$ 
is defined to be a vector satisfying 
\begin{align}\label{eq:intro:dG}
T_1\ket{G;q,t}=\Lambda^2\ket{G;q,t},\quad
T_n\ket{G;q,t}=0\ (n\ge2),
\end{align}
where $\Lambda^2$ is a (non-zero) complex number.
The dual vector $\bra{G;q,t}\in \widehat{M}_h^*$ is defined similarly:
\begin{align*}
\bra{G;q,t}T_{-1}=\Lambda^2\bra{G;q,t},\quad
\bra{G;q,t}T_n=0\ (n\le-2).
\end{align*}

\subsubsection{Recollection of the $K$-theoretic Nekrasov partition function}
\label{sec:intro:subsec:AGT:subsubsec:Nek}

The instanton part of the $\SU(N)$ $K$-theoretic Nekrasov partition function  
was defined as the integration in the equivariant $K$-theory on the 
moduli space of framed rank $N$ torsion free sheaves on $\bbP^2$ 
in \cite{N:2003} (see also \cite{NY:2005:b}). 
By the localization theorem for equivariant $K$-theory, 
it becomes a summation over the fixed point contributions. 
The fixed points are parametrised by $N$-tuples of Young diagrams, 
and one obtains a combinatorial form of the partition function.

Recall the definition of the Nekrasov factor $N_{\lambda,\mu}(u)$ in 
Definition \ref{dfn:N}.
The Nekrasov partition function for the $\SU(2)$ case is given as follows:
\begin{align*}
&Z_{\text{pure} \SU(2)}^{\text{inst} K}(Q,q,t;\Lambda)
=\sum_{\lambda,\mu\in\calP}
 (\Lambda^4 t/q)^{|\lambda|+|\mu|} Z_{\lambda, \mu}(Q,q,t),
\\
&Z_{\lambda,\mu}(Q,q,t) \seteq
 \Bigl[N_{\lambda,\lambda}(1)N_{\mu,\mu}(1)
       N_{\lambda,\mu}(Q)N_{\mu,\lambda}(Q^{-1})\Bigr]^{-1}.
\end{align*}

\subsubsection{$K$-theoretic AGT conjecture}\label{sss:KAGT}
Now the main conjecture in \cite{AY:2010} is:
under the correspondence $h=Q^{1/2}+Q^{-1/2}$ we have 
\begin{align}\label{eq:KAGT}
\langle G;q,t | G;q,t \rangle 
\stackrel{?}{=} Z_{\text{pure} \SU(2)}^{\text{inst} K}(Q,q,t;\Lambda).
\end{align}
Here the pairing of $M_h$ and $M_h^*$ is determined 
by $\langle h | h \rangle=1$.

\cite{AY:2010} also proposes a factorized expansion of $\ket{G;q,t}$ 
in terms of the Macdonald symmetric functions:
under certain identification of $M_h$ 
and the Fock space $\calF\simeq\Lambda_{\bbF}$, 
\begin{align}\label{eq:intro:dGP}
\ket{G;q,t}\stackrel{?}{=}
\sum_{\lambda}
 \Lambda^{2|\lambda|} P_{\lambda}(x;q,t)
 \prod_{\square\in\lambda}
 \dfrac{Q^{1/2}}{1-Q q^{i(\square)} t^{-j(\square)}}
 \dfrac{q^{a_\lambda(\square)}}
       {1-q^{a_\lambda(\square)+1}t^{\ell_\lambda(\square)}}.
\end{align}
Here the index $\lambda$ runs over the set of arbitrary partitions.
Note that the four dimensional version of this expansion is 
proved in \cite{Y:2011}.

A remark is in order here. In the expansion \eqref{eq:intro:dGP}, the 
pairing in \eqref{eq:KAGT} is not consistent with the pairing on $\calF$ 
induced from the Macdonald inner product \eqref{eq:Macdpair} 
on $\Lambda_{\bbF}$.
The former expressed in terms of power-sum symmetric function is 
$\langle p_m, p_n \rangle=-n(t^n+q^n)(q/t)^n \tfrac{1-q^n}{1-t^n}\delta_{m,n}$ 
(see \cite[(3.20)]{AY:2010}), 
but the latter is 
$\langle p_m, p_n \rangle_{q,t}=n\tfrac{1-q^n}{1-t^n}\delta_{m,n}$.

%%%%%%%%%%%%%%%%%%%%%%%%%%%%%%%%%%%%%%%%%%%%%%%%%%
%%%%%%%%%%%%%%%%%%%%%%%%%%%%%%%%%%%%%%%%%%%%%%%%%%

\section{Whittaker vectors for Ding-Iohara algebra}
\label{sect:whit}

The purpose of this section is the introduction of the 
Whittaker vector for the level $m$ representation of $\calU$.
First we give a construction of the Whittaker vector $\ket{G}$
for the level one representation, 
and show that the expansion of $\ket{G}$ 
in terms of the Macdonald symmetric function has factorized coefficients.
Next we define its higher level analogue $\ket{G;\Lambda,\bfal}$.
We give a conjecture on the factorized coefficients of $\ket{G;\Lambda,\bfal}$ 
expanded in the basis $(\ket{P_\vlambda})$.

\subsection{Level one case}

We introduce the Whittaker (or Gaiotto) vector
for the level one representation of the Ding-Iohara algebra.
Our argument here is based on Macdonald's  
homomorphism $\varepsilon_{u,t} \Lambda_\bbF \rightarrow \bbF$ 
defined by $\varepsilon_{u,t} (p_r)=(1-u^r)/(1-t^r)$ 
 \cite[Chap. VI, (6.16)]{M:1995}, and the 
 factorization formula for $\varepsilon_{u,t} (P_\lambda)$ \cite[Chap. VI, (6.17)]{M:1995}.
%%%%%%%%%%%%%%%
\subsubsection{Whittaker vector $\ket{G}$}

For two parameters $\alpha,\beta$, let $\ket{G}$ be the vector
\begin{align} \label{eq:Whit:G}
\ket{G}\seteq
 \exp\Big(
    \sum_{n=0}^\infty \dfrac{1}{n}\dfrac{\beta^n-\alpha^n}{1-q^n}a_{-n}
    \Big)\ket{0},
\end{align}
in the completed Fock space $\widehat{\calF}_u$.
{}From the permutation relation
\begin{align*}
\eta(z) \exp\Big(
    \sum_{n=0}^\infty \dfrac{1}{n}\dfrac{\beta^n-\alpha^n}{1-q^n}a_{-n}
    \Big)
=\dfrac{1-\beta/z}{1-\alpha/z} \exp\Big(
    \sum_{n=0}^\infty \dfrac{1}{n}\dfrac{\beta^n-\alpha^n}{1-q^n}a_{-n}
    \Big)
\eta(z),
\end{align*}
we have
\begin{align*}
(1-\alpha/z)\eta(z)\ket{G}
=(1-\beta/z)  \exp\Big(
    \sum_{n=0}^\infty \dfrac{1}{n}\dfrac{\beta^n-\alpha^n}{1-q^n}a_{-n}
    \Big)\eta(z)\ket{0}.
\end{align*}
In Fourier modes, we have
\begin{align}\label{eq:level1:whit:cond}
(\eta_{n+1}-\alpha\eta_n) \ket{G}=
\begin{cases}
-\beta \ket{G} &n=0,            \\
0              &n=1,2,3,\ldots.
\end{cases}
\end{align}
Therefore $\ket{G}$ is a joint eigenfunction 
with respect to the set of operators $\eta_{n+1}-\alpha\eta_n$
$(n=0,1,2,\cdots)$.
Note that these equations resemble the defining conditions of Whittaker vectors 
(see \eqref{eq:G:intro} and \eqref{eq:intro:dG}).

Note that by setting $\alpha=v$ and $\beta=tu/q$, we have
$\ket{G}=\Phi(w)\ket{0}$, where $\Phi(w)$ is the vertex operator in Proposition \ref{Phi=}.
Then (\ref{eq:level1:whit:cond}) can be regarded as a 
good starting point for guessing the 
correct permutation relations for $\Phi(w)$ stated in 
Definition \ref{defVOlevelone}.

%%%%%%%%%%%%%%%%%%%%%%%%%%%%%%
\subsubsection{Factorized coefficients of Whittaker vector}

We show that the expansion of $\ket{G}$  
in terms of $P_\lambda$ has factorized coefficients.
One may compare this with  \eqref{eq:intro:dGP}. 
Note that, however, here we treat the level one case and 
\eqref{eq:intro:dGP} is related with the 
deformed Virasoro algebra i.e. the level two representation from 
the point of view of $\calU$.

Recall the following specialization 
of $P_\lambda(x;q,t)$ \cite[VI, (6.17)]{M:1995}.
Let $u\in \bbF$.
Then under the homomorphism
\begin{align*}
\ve_{u,t}:\Lambda_{\bbF} \to \bbF, \quad p_n\mapsto \dfrac{1-u^n}{1-t^n},
\end{align*}
we have
\begin{align}\label{eq:level1:specP}
\ve_{u,t} P_\lambda = 
\prod_{\square\in\lambda} 
 \dfrac{t^{i(\square)-1}-q^{j(\square)-1}u}
 {1-q^{a_\lambda(\square)}t^{\ell_\lambda(\square)+1}}.
\end{align}
For simplicity of display, let 
\begin{align*}
\widetilde{\ve}_{\alpha,\beta,t}:
\Lambda_{\bbF} \to \bbF, \quad p_n\mapsto \dfrac{\beta^n-\alpha^n}{1-t^n}.
\end{align*}
From \eqref{eq:level1:specP} and \eqref{eq:level1:J}, we have
\begin{align}\label{eq:level1:specQ}
\widetilde{\ve}_{\alpha,\beta,t} Q_\lambda 
 =\beta^{|\lambda|} \cdot \ve_{\alpha/\beta,t} Q_\lambda 
 =(\beta^{|\lambda|}c_\lambda/c_\lambda') \cdot \ve_{\alpha/\beta,t} P_\lambda
 =\prod_{\square\in\lambda} 
  \dfrac{t^{i(\square)-1}\beta-q^{j(\square)-1}\alpha}
        {1-q^{a_\lambda(\square)+1}t^{\ell_\lambda(\square)}}.
\end{align}

Next we recall the Cauchy-type 
kernel function \cite[VI \S 2, (4.13)]{M:1995}:
\begin{align}\label{eq:level1:kernel}
\exp\Big(\sum_{n=1}^\infty \dfrac{1}{n}\dfrac{1-t^n}{1-q^n}p_n(x)p_n(y)\Big)
=\sum_{\lambda} P_\lambda(x;q,t)Q_\lambda(y;q,t),
\end{align}
where the index $\lambda$ runs over the set of arbitrary partitions.

Now let us apply the specialization $\widetilde{\ve}_{\alpha,\beta,t}$
to the $y$-variables in \eqref{eq:level1:kernel}.
We have
\begin{align*}
\exp\Big(
 \sum_{n=1}^\infty \dfrac{1}{n}\dfrac{\beta^n-\alpha^n}{1-q^n}p_n(x)\Big)
=\sum_{\lambda} 
  P_\lambda(x;q,t) 
  \cdot \prod_{\square\in\lambda} 
  \dfrac{t^{i(\square)-1}\beta -q^{j(\square)-1}\alpha}
        {1-q^{a_\lambda(\square)+1}t^{\ell_\lambda(\square)}}.
\end{align*}
Since the left hand side is $\ket{G}$ 
under the identification $\calF_u\simto\Lambda_{\widetilde{\bbF}}$ 
\eqref{eq:intro:iota}, we have
\begin{prp}
We have \begin{align*}
\ket{G}=\sum_{\lambda} 
 \ket{ P_\lambda(x;q,t) }
  \cdot \prod_{\square\in\lambda} 
  \dfrac{t^{i(\square)-1}\beta -q^{j(\square)-1}\alpha}
        {1-q^{a_\lambda(\square)+1}t^{\ell_\lambda(\square)}}.
\end{align*}
\end{prp}

\subsection{Conjecture for the Higher level cases}

For the higher level representations, 
we introduce the Whittaker vector as follows.

\begin{dfn}
For an $m$-tuple of parameters $\bfal=(\alpha_1,\ldots,\alpha_m)$ and 
a parameter $\Lambda$,
let us define the state $\ket{G;\Lambda,\bfal}\in\widehat{\calF}_{\bfu}$ 
of the completed Fock space by the condition
\begin{align}\label{eq:levelm:whit:cond}
\left(
% (q/t)^{(1-k)/2}
X^{(k)}_n-\Lambda_{\bfal}X^{(k)}_{n-1}
 +e_k(\bfal/v)\Lambda_{\bfal}\delta_{n,1}
\right)\ket{G;\Lambda,\bfal}=0 
\quad
(n\in\bbZ_{\ge1},\ k\in\bbZ_{\ge0}),
\end{align}
with the normalization condition
\begin{align*}
\ket{G;\Lambda,\bfal}=\ket{\bf 0}+\cdots.
\end{align*}
Here we used the symbols
\begin{align*}
v\seteq \sqrt{q/t},\quad
\Lambda_{\bfal} \seteq \Lambda \prod_{i=1}^m v u_i/\alpha_i,
\end{align*}
and $e_k(\bfal/v)$ is the $k$-th elementary symmetric polynomial 
in $(\alpha_1/v,\alpha_2/v,\ldots,\alpha_m/v)$.
\end{dfn}
%This is an analogue of \eqref{eq:level1:whit:cond}.
%Note that
%$ {\DITn i{0}}\Gaiotto=e_i(\vQ)\Gaiotto$.

The condition \eqref{eq:levelm:whit:cond} can be rewritten 
as follows.
Let 
\begin{align*}
\ket{G;\Lambda,\bfal}= 
\sum_{n=0}^{\infty}\Lambda_{\bfal}^n \ket{G;\Lambda,\bfal;n},
\end{align*}
be the expansion of $\ket{G;\Lambda,\bfal}$ 
with  $\ket{G;\Lambda,\bfal;n}\in\calF_{\bfu,n}$.
Here $\calF_{\bfu,n}$ is the homogeneous component of $\calF_{\bfu}$ 
whose degree is induced by that of $\calF=\oplus_n \calF_n$.
In other words, 
$\calF_{\bfu,n}\seteq
 \oplus_{n_1+\cdots+n_m=n}\calF_{n_1}\otimes\cdots\otimes\calF_{n_m}$.
Then \eqref{eq:levelm:whit:cond} is equivalent to
\begin{align*}
%(q/t)^{(1-k)/2}
X^{(k)}_n \ket{G;\Lambda,\bfal;n}
- X^{(k)}_{n-1} \ket{G;\Lambda,\bfal;n-1} 
+ e_k(\bfal/v) \delta_{n,1} \ket{G;\Lambda,\bfal;n-1} 
=0.
\end{align*}

The dual state is defined as follows.
For an $m$-tuple of parameters $\bfbe=(\beta_1,\ldots,\beta_m)$, 
let $\bra{G;\Lambda,\bfbe}\in\widehat{\calF}_{\bfu}^*$ be 
an element such that
\begin{align*}
\bra{G;\Lambda,\bfbe}\left(
 (q/t)^{1-k} X_{-n}^{(k)}-\Lambda_{\bfbe} X_{1-n}^{(k)} 
 + e_i(\bfbe/v) \Lambda_{\bfbe} \delta_{n,1}\right)=0
\end{align*}
with
\begin{align*}
\Lambda_{\bfbe} \seteq \Lambda \prod_{i=1}^m v u_i/\beta_i
\end{align*} 
and the normalization condition 
$\bra{G;\Lambda,\bfbe}=\Vec{1}+\cdots$.

Now we state our conjecture.
\begin{con}
(1)
For generic parameters, $\ket{G;\Lambda,\bfal}$ exists uniquely.
We have the expansion 
\begin{align*}
&\ket{G;\Lambda,\bfal}=
\sum_\vlambda (q/t)^{ \sum_{k=1}^m \frac{1-k}{2} | \lambda^{(k)} | }
C_{\vlambda}(\Lambda,\bfu,\bfal; q,t) \ket{P_\vlambda},
\end{align*}
where
\begin{align*}
C_{\vlambda}(\Lambda,\bfu,\bfal; q,t) \stackrel{?}{=}
&\prod_{k=1}^m \Bigl[
 \prod_{l=k+1}^{m} (u_k/u_l)^{|\lambda^{(l)}|}
 \times
 \prod_{\square \in \lambda^{(k)} } \Bigl[
 \dfrac{-\Lambda q^{-j(\square)} (-q^{1-j(\square)}t^{i(\square)-1})^{k-1}}
      {1-q^{-a_{\lambda^{(k)}}(\square)-1}t^{-\ell_{\lambda^{(k)}}(\square)}}
\\
&\quad \times
 \dfrac{\displaystyle 
        \prod_{l=1}^{m}
        \Bigl(1-v\dfrac{u_k}{\alpha_l}q^{j(\square)-1}t^{1-i(\square)}\Bigr)}
       {\displaystyle \prod_{l=1}^{k-1}
        \Bigl(1-\dfrac{u_l}{u_k} q^{a_{\lambda^{(l)}}(\square)} 
              t^{\ell_{\lambda^{(k)}}(\square)+1} \Bigr)
        \prod_{l=k+1}^{m}
        \Bigl(1-\dfrac{u_k}{u_l} q^{-a_{\lambda^{(l)}}(\square)-1}
              t^{-\ell_{\lambda^{(k)}}(\square)} \Bigr)}\Bigr]\Bigr]
\\
=\Bigl[\prod_{k=1}^m \prod_{\square \in \lambda^{(k)} } 
 &\dfrac{-\Lambda q^{-j(\square)} (-q^{1-j(\square)}t^{i(\square)-1})^{k-1}}
        {1-q^{-a_{\lambda^{(k)}}(\square)-1}t^{-\ell_{\lambda^{(k)}}(\square)}}
  \Bigr]
  \cdot
  \Bigl[\dfrac{%\displaystyle
               \prod_{k,l=1}^{m} N_{\lambda^{(k)},\emptyset}(v u_k/\alpha_l)}
              {%\displaystyle
               \prod_{1\le k < l \le m} 
               \bigl(\dfrac{u_k}{u_l}\bigr)^{-|\lambda^{(l)}|}
               N_{\lambda^{(k)},\lambda^{(l)}}(u_k/u_l)}
  \Bigr].
\end{align*}

(2)
For generic parameters, the element $\bra{G;\Lambda,\bfbe}$ exists uniquely.
We have
\begin{align*} 
\bra{G;\Lambda,\bfbe}=
&\sum_{\vlambda}
(q/t)^{ \sum_{k=1}^m \frac{k-1}{2} | \lambda^{(k)} | }
\overline{C}_{\vlambda}(\Lambda,\bfu,\bfbe;q,t)
 \bra{P_{\vlambda}},
\end{align*}
where
\begin{align*}
\overline{C}_{\vlambda}(\Lambda,\bfu,\bfbe;q,t)
\stackrel{?}{=}
&C_{\overline{\vlambda}}(\Lambda,\overline{\bfu},\bfbe;q,t)
\stackrel{?}{=}
C_{\overline{\vlambda}}
 (\Lambda_{\bfbe},1/\overline{\bfu},v^2/\bfbe;q^{-1},t^{-1})
\\
\stackrel{?}{=}
 \Bigl[\prod_{k=1}^m \prod_{\square \in \lambda^{(k)} } 
 &\dfrac{-\Lambda q^{-j(\square)} (-q^{1-j(\square)}t^{i(\square)-1})^{m-k}}
        {1-q^{-a_{\lambda^{(k)}}(\square)-1}t^{-\ell_{\lambda^{(k)}}(\square)}}
  \Bigr]
  \cdot
  \Bigl[\dfrac{%\displaystyle
               \prod_{k,l=1}^{m} N_{\lambda^{(k)},\emptyset}(v u_k/\beta_l)}
              {%\displaystyle
               \prod_{1\le k < l \le m} 
               \bigl(\dfrac{u_l}{u_k}\bigr)^{-|\lambda^{(k)}|}
               N_{\lambda^{(l)},\lambda^{(k)}}(u_l/u_k)}
  \Bigr]
\\
=\Bigl[\prod_{k=1}^m \prod_{\square \in \lambda^{(k)} } 
 &\dfrac{-\Lambda_{\bfbe} q^{j(\square)-1} 
         (-q^{1-j(\square)}t^{i(\square)-1})^{1-k}}
        {1-q^{a_{\lambda^{(k)}}(\square)+1}t^{\ell_{\lambda^{(k)}}(\square)}}
  \Bigr]
  \cdot
  \Bigl[\dfrac{%\displaystyle
               \prod_{k,l=1}^{m} N_{\emptyset,\lambda^{(k)}}(v \beta_l/u_k)}
              {%\displaystyle
               \prod_{1\le k < l \le m} 
               \bigl(\dfrac{u_l}{u_k}\bigr)^{-|\lambda^{(l)}|}
               N_{\lambda^{(l)},\lambda^{(k)}}(u_l/u_k)}
  \Bigr].
\end{align*}
Here we used 
$\overline{\vlambda}\seteq(\lambda^{(m)},\ldots,\lambda^{(2)},\lambda^{(1)})$,
$\overline{\bfu} \seteq (u_m,\ldots,u_2,u_1)$,
$1/\overline{\bfu} \seteq (1/u_m,\ldots,1/u_2,1/u_1)$, 
and $v^2/\bfbe \seteq (v^2/\beta_1,\ldots,v^2/\beta_m)$.
\end{con}

This conjecture implies 
\begin{align*}%\label{eq:Whit:norm}
\langle G;\Lambda,\bfbe | G;\Lambda,\bfal \rangle =
&\sum_{\vlambda}
 C_{\vlambda}(\Lambda,\bfu,\bfal; q,t) 
 \overline{C}_{\vlambda}(\Lambda,\bfu,\bfbe;q,t)
 \prod_{k=1}^m \dfrac{c'_{\lambda^{(k)}}}{c_{\lambda^{(k)}}} 
\\
\stackrel{?}{=}
&\sum_{\vlambda} 
%\biggl(\Lambda^2 v^m \dfrac{e_m(\bfu)}{e_m(\bfbe)}\biggr)^{|\vlambda|} 
 \bigl(\Lambda \Lambda_{\bfbe}\bigr)^{|\vlambda|} 
 \prod_{k,l=1}^{m} 
 \dfrac{N_{\lambda^{(k)},\emptyset}(v u_k/\alpha_l) 
        N_{\emptyset,\lambda^{(k)}}(v \beta_l/u_k)}
       {N_{\lambda^{(k)},\lambda^{(l)}}(u_k/u_l)}.
\end{align*}
This is equal to the instanton part of the five dimensional 
$\U(m)$ Nekrasov partition function with $N_f=2m$ fundamental 
matters (see \cite[\S 5]{AY:2010:2}).

%%%%%%%%%%%%%%%%%%%%%%%%%%%%%%%%%%%%%%%%%%%%%%%%%%
%%%%%%%%%%%%%%%%%%%%%%%%%%%%%%%%%%%%%%%%%%%%%%%%%%

\section{Examples of the
Matrix elements of the Level One Vertex Operator}
\label{sect:vertex}
{}From the point of view of the Whittaker vector considered 
in the last section
(in particular the expression \eqref{eq:Whit:G}),
one may be interested in the operator 
\begin{align}
\psi(z)=\psi(z;\alpha,\beta,\kappa,\delta):= \exp \Bigl(
 \sum_{n=1}^{\infty} \dfrac{1}{n}\dfrac{\beta^n-\alpha^n}{1-q^n} a_{-n}w^n 
      \Bigr)
 \exp\Bigl(
  \sum_{n=1}^{\infty} 
  \dfrac{1}{n}\dfrac{\delta^n- \kappa^n}{1-q^{n}} a_{n}w^{-n}
     \Bigr),
\end{align}
where $\alpha$, $\beta$, $\kappa$ and $\delta$ are parameters.
However, one may easily find that we need to have
some relations in the parameters $\alpha$, $\beta$, $\kappa$ and $\delta$ if 
we demand that all the matrix elements 
with respect to the Macdonald functions be factorized.

In this section, we give some examples of the calculation of the 
matrix elements, in which some transformation formulas for 
the basic hypergeometric series can be applied.
Recall \eqref{eq:level1:Phi}, where we have 
$\Phi(w):\calF_u\to\calF_v$ and 
\begin{align}
&\Phi(w) =
 \exp \Bigl(
 -\sum_{n=1}^{\infty} \dfrac{1}{n}\dfrac{v^n-(t/q)^{n}u^n}{1-q^n} a_{-n}w^n 
      \Bigr)
 \exp\Bigl(-
  \sum_{n=1}^{\infty} \dfrac{1}{n}\dfrac{q^nv^{-n}- q^n u^{-n}}{1-q^{n}} a_{n}w^{-n}
     \Bigr).
\end{align}
Namely, we have $\Phi(w) =\psi(w;v,t u/q,q/v,q/u) $.
We use the notations
\begin{align*}
&(x;q)_n\seteq\prod_{i=1}^n(1-x q^{i-1}),\quad
(x_1,x_2,\ldots,x_m;q)_n\seteq \prod_{i=1}^m(x_i;q)_n,\\
&
{}_3\phi_2(a_1,a_2,a_3;b_1,b_2;q,z)
=
\sum_{n=0}^{\infty}
 \dfrac{(a_1;q)_n(a_2;q)_n(a_3,q)_n}{(q;q)_n(b_1;q)_n(b_2;q)_n}z^n.
\end{align*}

%%%%%%%%%%%%%%%%%%%%%%%%%%%%%%
\subsection{The case $\lambda=(j)$, $\mu=(k)$}
We will compute $\bra{ Q_{(j)}} \psi(w) \ket{ Q_{(k)}}$ 
for $j,k\in\bbZ_{\ge0}$.

We have the generating function of $Q_{(r)}$
\begin{align*}%\label{eq:level1:Qgen}
\sum_{r=0}^\infty Q_{(r)}(x;q,t) y^r
=\exp\Big(\sum_{n=1}^\infty \dfrac{1}{n}\dfrac{1-t^n}{1-q^n}p_n(x)y^n\Big).\nonumber
\end{align*}
Using the isomorphism $\iota: \Lambda_{\FF}\simto\calF$ \eqref{eq:intro:iota} 
we have
\begin{align*}
\nonumber
&
 \sum_{j,k=0}^\infty 
 \bra{Q_{(j)}} \psi(w) \ket{Q_{(k)}}
 x^{-j} y^{k}
\\
&\nonumber
=\Bigl\langle 
   \exp\Big(\sum_{k=1}^\infty \dfrac{1}{k}\dfrac{1-t^k}{1-q^k}a_k x^{-k}\Big)
   \psi(w)
   \exp\Big(\sum_{k=1}^\infty \dfrac{1}{k}\dfrac{1-t^k}{1-q^k}a_{-k} y^k\Big)
  \Bigr\rangle\\
  &
 \nonumber% \label{eq:level1:yy:1}
  =\dfrac{(\alpha w/x  ;q)_\infty}{(\beta w/ x;q)_\infty}
  \dfrac{(t y/x  ;q)_\infty}{(y    /x  ;q)_\infty}
  \dfrac{(\kappa y/ w;q)_\infty}{(\delta  y/ w;q)_\infty}\\
  &
  =\sum_{l,m,n=0}^\infty
   w^{l-n}x^{-l-m}y^{m+n}
   \dfrac{(\alpha/\beta;q)_l (t;q)_m (\kappa/\delta;q)_n}{(q;q)_l (q;q)_m (q;q)_n}
\beta^{l}
 \delta^{n}. 
\end{align*}
Here we have used the $q$-binomial formula 
$(a z;q)_\infty /(z;q)_\infty = \sum_{n=0}^\infty z^n (a;q)_n/(q;q)_n$.
Hence we have
\begin{align}
& \bra{Q_{(j)}} \psi(w) \ket{Q_{(k)}}=
  w^{j-k}\beta^{j}
 \delta^{k}
   \dfrac{(\alpha/\beta;q)_{j} (\kappa/\delta;q)_{k}}{(q;q)_{j} (q;q)_{k}}\nonumber\\
   &\qquad \times
\sum_{m=0}^{\infty}
   \dfrac{(q^{-j};q)_{m} (t;q)_m (q^{-k};q)_{m}}
   {(q^{-j+1}\beta/\alpha;q)_{m} (q;q)_m (q^{-k+1}\delta/\kappa;q)_{m}}
   q^{2m}\alpha^{-m}\kappa^{-m}.\label{QQ}
   \end{align}

Recall the $q$-analogue of Saalsch\"{u}tz's summation 
formula for terminating balanced ${}_3\phi_2$ series \cite[\S 1.7]{GR:2004}:
\begin{align}\label{eq:level_m:saal}
{}_3\phi_2(a,b,q^{-k};c,ab c^{-1}q^{1-k};q,q)
=\dfrac{(c/a,c/b;q)_k}{(c,c/ab;q)_k}.
\end{align}
Let 
\begin{align*}
a=q^{-j},\qquad b=t,\qquad c=q^{-j+1}\beta/\alpha.
\end{align*}
Then we have the two conditions
\begin{align} \label{comp}
\alpha\kappa=q ,\qquad \beta\delta=t
\end{align}
to identify the two ${}_3\phi_2$ series 
in \eqref{QQ} and \eqref{eq:level_m:saal}.
Setting $\alpha=v$ and $\beta=tu/q$, we have $\kappa=q/v$ and $\delta=q/u$ 
{}from \eqref{comp}.
Thus we conclude that we have factorized matrix elements with respect to 
the one row Macdonald function $Q_{(n)}$'s for the operator
 $\Phi(w) =\psi(w;v,t u/q,q/v,q/u) $.

Noting that $c_{(n)}'=(q;q)_n$ and simplifying the formulas, we have
\begin{align*}
\bra{J_{(j)}} \Phi(w) \ket{J_{(k)}}
=&(q^{-k+1} v/t u;q)_{j} (q^{1+j-k} v/u;q)_{k}
  w^{j-k}(t u/q)^{j}(-v/q)^{-k} q^{k(k-1)/2},
\end{align*}
which agrees with Proposition \ref{conj:level1}.

%%%%%%%%%%%%%%%%%%%%%%%%%%%%%%%
\subsection{The case $\lambda=(1^j)$, $\mu=(k)$}
Next we treat the case when the partition $\lambda$ is one column 
and $\mu$ is one row.

In this case we have $J_{(1^j)}=c_{(1^j)} P_{(1^j)}=c_{(1^j)} e_{j}$,
where $e_j$ is the $j$-th elementary symmetric function.
The generating function is given by 
\begin{align*}
\sum_{r=0}^{\infty} e_{r}(x) y^r
=\exp\Big(-\sum_{n=1}^\infty \dfrac{1}{n} p_n(x)(-y)^n\Big).
\end{align*}
We have
\begin{align*}
\nonumber
&
 \sum_{j,k=0}^\infty 
 \bra{ P_{(1^j)}} \Phi(w) \ket{Q_{(k)}}
 x^{-j} y^{k}
\\
&=\Bigl\langle 
 \exp\Big(-\sum_{k=1}^\infty \dfrac{1}{k}a_k (-x)^{-k}\Big)
 \Phi(w)
 \exp\Big(\sum_{k=1}^\infty \dfrac{1}{k}\dfrac{1-t^k}{1-q^k}a_{-k} y^k\Big)
 \Bigr\rangle\\
 &=\dfrac{(-t w u/qx ;t )_\infty}{(-v w/x;t)_\infty}
  \Bigl(1+\dfrac{y}{x}\Bigr)
  \dfrac{(qy/v w;q)_\infty}{(qy/u w;q)_\infty}\\
  &
  =\Bigl[\sum_{m=0}^\infty
        \dfrac{(t u/q v;t)_m}{(t;t)_m}\Big(-\dfrac{v w}{x}\Big)^m\Bigr]
  \cdot
  \Bigl[1+\dfrac{y}{x}\Bigr]
  \cdot
  \Bigl[\sum_{n=0}^\infty \dfrac{(u/v;q)_n}{(q;q)_n}
        \Bigl(\dfrac{qy}{u w}\Bigr)^n\Bigr].
\end{align*}
Hence we have
\begin{align*}
 &\bra{ P_{(1^j)}} \Phi(w) \ket{Q_{(k)}}=(-vw)^j (q/uw)^k
   \dfrac{(t u/q v;t)_j}{(t;t)_j}
       \dfrac{(u/v;q)_k}{(q;q)_k}
   {(1-u/qv)(1-q^{k-1}t^ju/v)\over (1-q^{k-1}u/v)(1-t^ju/qv)}.
\end{align*}

Recalling $c_{(1^j)}=(t;t)_j$ and $c'_{(k)}=(q;q)_k$, we have
\begin{align*}
\bra{J_{(1^j)}} \Phi(w) \ket{J_{(k)}}
&=(1-q^{1-k}t^{-j}v/u) 
(q t^{1-j}v/u;t)_{j-1}
  (q^{2-k}v/u;q)_k
\\
&\quad
  \times w^{j-k}(t u/q)^j (-v/q)^{-k}t^{j(j-1)/2}q^{k(k-1)/2}.
\end{align*}

%%%%%%%%%%%%%%%%%%%%%%%%%%%%
\subsection{The case $\lambda=(1^j)$, $\mu=(1^k)$}
This case is similar to the first case $\lambda=(j)$, $\mu=(k)$.
The generating function we consider is
\begin{align*}
\nonumber
&\sum_{j,k=0}^\infty 
 \bra{P_{(1^j)}} \Phi(w) \ket{P_{(1^k)}}
x^{-j} y^{k}
\\
&=\Bigl\langle 
 \exp\Big(-\sum_{k=1}^\infty \dfrac{1}{k}a_k (-x)^{-k}\Big)
 \Phi(w)
 \exp\Big(-\sum_{k=1}^\infty \dfrac{1}{k} a_{-k} (-y)^k\Big)
 \Bigr\rangle\\
 &
 =\dfrac{(-t u w/q x;t)_\infty}{(-v w/ x;t)_\infty}
 \dfrac{(q y/x;t)_\infty}{(y/x;t)_\infty}
 \dfrac{(-q y/u w;t)_\infty}{(-q y/v w;t)_\infty}.
\end{align*}
Then we have
\begin{align*}
&\bra{P_{(1^j)}} \Phi(w) \ket{P_{(1^k)}}=
(-vw)^j(-q/vw)^k {(tu/qv;t)_j\over (t;t)_j}{(v/u;t)_k\over (t;t)_k}\\
&\qquad \times \sum_{m=0}^{\infty}
{(q;t)_m(t^{-j};t)_m(t^{-k},t)_m \over 
(t;t)_m (t^{-j}qv/u;t)_m (t^{-k+1}u/v;t)_m}t^m.
\end{align*}
Using the $q$-Saalsch\"{u}tz's formula \eqref{eq:level_m:saal},
we have
\begin{align*}
 \bra{ J_{(1^j)}} \Phi(w) \ket{J_{(1^k)}}
=&(t^{-j}v/u;t)_{k} (q t^{-j+k}v/u;t)_{j} 
\\
 &\times
 w^{j-k} (t u/q)^{j} (-v/q)^{-k} t^{j(j-1)/2}.
\end{align*}

%%%%%%%%%%%%%%%%%%%%%%%%%%%%%%%%%%%%%%%%%%%%%%%%%%
%%%%%%%%%%%%%%%%%%%%%%%%%%%%%%%%%%%%%%%%%%%%%%%%%%

\end{document}